\documentclass[prr,letterpaper,aps,10pt,superscriptaddress,twocolumn,floatfix,showpacs]{revtex4-2}
\usepackage{amsmath, graphicx, bm, amsfonts, hyperref, placeins, xcolor, dsfont, physics, algorithm, algpseudocode, booktabs, mathtools, siunitx}
\usepackage[expansion=true,protrusion=true]{microtype}

\hypersetup{breaklinks=true}
\usepackage[noabbrev, capitalise, nameinlink]{cleveref}
\crefname{section}{Sect.}{Sects.}
\crefname{figure}{Fig.}{Figs.}
\crefname{equation}{Eq.}{Eqs.}
\crefname{appendix}{App.}{Apps.}

\newcommand*\diff{\mathop{}\!\mathrm{d}}

\newcommand{\me}{\mathrm{e}}
\newcommand{\iu}{\mathrm{i}}

\newcommand{\Lm}{\mathcal{L}}
\newcommand{\Hm}{\mathcal{H}}
\newcommand{\Nm}{\mathcal{N}}

\newcommand{\Ne}{\mathcal{N}^{(e)}}
\newcommand{\Ng}{\mathcal{N}^{(g)}}
\newcommand{\gd}[1]{\hat{g}^\dag_{#1}}
\newcommand{\g}[1]{\hat{g}^{\phantom\dag}_{#1}}
\newcommand{\ed}[1]{\hat{e}^\dag_{#1}}
\newcommand{\e}[1]{\hat{e}^{\phantom\dag}_{#1}}

\NewDocumentCommand{\Sig}{ m g }{%
  \hat{\Sigma}^{\IfNoValueTF{#2}{\vphantom\dag}{#2}}_{#1}%
}

\newcommand{\Sigd}[1]{\hat{\Sigma}^{\dag}_{#1}}

\newcommand{\Gammalower}{\Tilde{\gamma}_\mathrm{L}}

\newcommand{\kb}{k_\mathrm{B}}
\binoppenalty=\maxdimen
\relpenalty=\maxdimen

\newcommand{\tf}{\nu}

\begin{document}

\newcommand{\fau}{Department of Physics, Friedrich-Alexander Universit\"at Erlangen-N\"urnberg (FAU), Staudtstra{\ss}e 7,  D-91058 Erlangen, Germany}
\newcommand{\mpl}{Max  Planck  Institute  for  the  Science  of  Light,  Staudtstraße  2,  D-91058  Erlangen,  Germany}
\newcommand{\tud}{TU Darmstadt, Institute for Applied Physics, Hochschulstraße 4A, D-64289 Darmstadt, Germany}

\begin{abstract}
Collective radiance effects in quantum degenerate systems, such as superradiance and subradiance of a partially inverted ensemble, are shaped by the interplay of spatial confinement and exchange statistics. We investigate this interplay using a purely dissipative field theoretic quartic Lindblad master equation, which captures the nonlinear dynamics of the combined motional and electronic manifolds. This approach captures the interplay between the permutational symmetry of the Lindbladian and the exchange symmetry of the particles, quantifying how bosonic enhancement and Pauli blocking dictate superradiant and subradiant scaling. We identify two distinct routes to distinguishable dynamics: thermal dilution of the initial state at high temperatures and the dynamical breakdown of collective order via recoil induced transport in soft traps. This analysis provides a benchmark for collective emission in quantum-degenerate atomic systems with coupled motional and internal dynamics, such as optical lattice clocks and spinor gases, when dissipation is engineered to control recoil and motional heating.
\end{abstract}

\title{Dicke superradiance in degenerate quantum matter: interplay of exchange statistics and spatial confinement}
\author{Julian Lyne}
\email{Julian.Lyne@fau.de}
\affiliation{\fau}
\affiliation{\mpl}
\author{Kai Phillip Schmidt}
\affiliation{\fau}
\author{Claudiu Genes}
\affiliation{\tud}
\author{Nico S. Bassler}
\email{nico.bassler@physik.tu-darmstadt.de}
\affiliation{\tud}
\maketitle

\section{Introduction}\label{sec:introduction}

\begin{figure*}[t]
    \centering
    \includegraphics[width=0.95\linewidth]{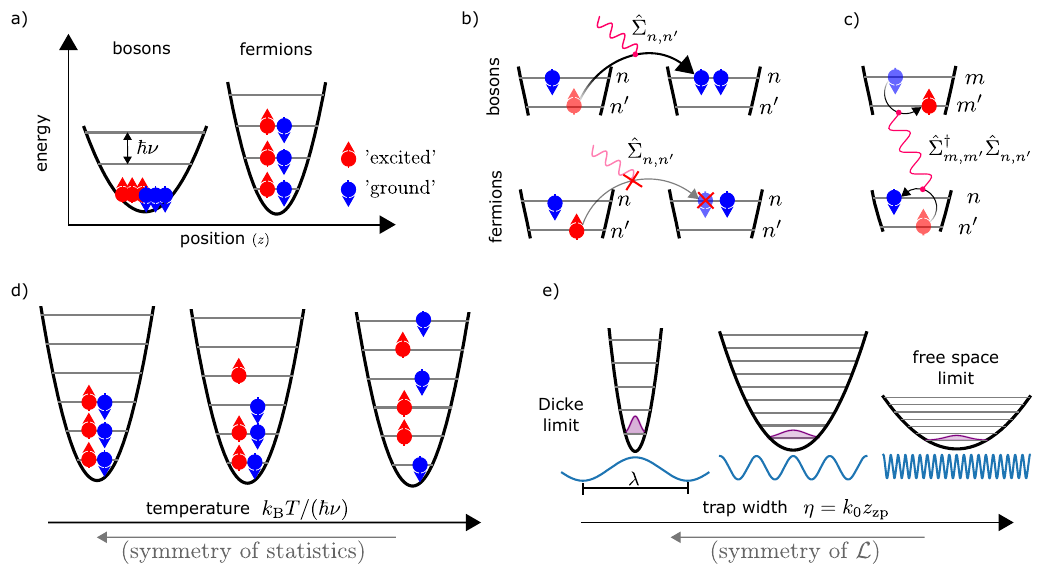}
    \caption{\textbf{Illustration of the model and symmetry competition in quantum gases: } a) Dual species degenerate Bose and Fermi gases in a 1-dimensional harmonic trap. b) Illustration of the influence of particle statistics on the dissipation process. While for bosons the decay of an excited particle into a level already occupied by a ground particle is enhanced, it is forbidden for fermions by the Pauli-exclusion principle. c) Illustration of the non-Hermitian Hamiltonian, where an excited particle and a ground particle exchange a photon, resulting in a change in the electronic and motional degrees of freedom. d) Symmetry of particle exchange diminishes as temperature is increased (for simplicity, only illustrated for fermions). e) Permutational symmetry of the Lindbladian decreases as the trap width is increased.}
    \label{fig:fig1}
\end{figure*}

Traditionally, collective radiance in quantum optical systems is investigated for a collection of distinguishable two level systems with a vast theoretical literature~\cite{Dicke54,bonifacio1971quantum1,bonifacio1971quantum2,Gross82,lee1977exact1,lee1977exact2,holzinger2025compact,holzinger2025solving,clemens2003collective,masson2022universality,sierra2022dicke,masson2024dicke,mok2024universal,holzinger2025symmetry,holzinger2025analyticalpredictionsuperradiantemission,sutherland2017superradiance,mok2025ground,holzinger2025symbolic} and experimental implementations across a broad variety of platforms~\cite{skribanowitz1973observation,gross1976observation,gibbs1977single,devoe1996observation,roof16,goban2015superradiance,solano2017super,bohnet2012steady,mlynek2014observation,scheibner2007superradiance,bradac2017room,pallmann2024cavity,rohlsberger2010collective,wang2007superradiance,okaba2019superradiance,ferioli2025emergence}. Collective radiance manifests in the form of super- and subradiance when modeling the dissipative dynamics by a Lindblad dissipator with collective collapse operators which are sums over the individual collapse operators. However, at high densities and low temperatures, emitters are fundamentally constrained by their exchange statistics, requiring a description in second quantization~\cite{lewenstein1994quantum,you1995quantum, you1996quantum}, where spontaneous emission is described as the annihilation of atoms of the excited type (referring to the electronic degree of freedom) and creation of atoms of the ground type. Although the effects of particle statistics on radiative properties have been studied extensively theoretically \cite{helmerson1990radiative,busch1998inhibition,osullivan2009spontaneous,gorlitz2001enhancement,shuve2009enhanced,sander2011spatial,sarkar2014light,bilitewski2022disentangeling}, only recently have experiments reached regimes where such effects can be observed directly~\cite{margalit2021pauli,demarco2021no,deb2021observation,sanner2021pauli,lu2023bosonic,jannin2022pauli}.

Here, we consider the radiative properties emerging from the interplay of collective radiance and particle statistics for quantum degenerate gases in an effectively 1-dimensional harmonic trap [see Fig.~\ref{fig:fig1}a),b),c)]. The problem is governed by two independent symmetry structures [see Fig.~\ref{fig:fig1}d),e)]. The first is the permutational symmetry of the Lindbladian, controlled by the spatial extent of the trap. Tight confinement yields the Dicke limit, while increasing trap width progressively weakens this symmetry. The second symmetry is the exchange symmetry of indistinguishable particles, imposed by particle statistics. This is not a symmetry in the sense of an invariance of the Lindbladian under some transformation, but rather fixes the algebraic relation of the operators appearing within the Lindbladian. Its dynamical relevance diminishes with increasing temperature as the gas becomes effectively dilute.

Previous theoretical treatments of radiative properties of quantum degenerate gases have largely focused on the low-excitation limit, where the field describing the excited internal state can be eliminated~\cite{cirac1994quantumPRA,cirac1994quantumPRL,javanainen1995Offresonance,ruostekoski1997quantum,ruostekoski1999opticalPRA,ruostekoski1999opticalPRL,ruostekoski2009light,cordobes2014optical,ruostekoski2016emergence}. In the quantum degenerate regime, the momentum transfer accompanied with optical transitions typically ejects particles from shallow traps, rendering such an approximation physically well justified. While particle statistics continue to constrain the accessible many-body states of the remaining degrees of freedom, eliminating the excited field removes explicit information about the internal-state dynamics and their coupling to motional degrees of freedom.

In this work, we consider a minimal model that retains both internal states dynamically and isolates the interplay between collective radiance and particle statistics. We study the purely dissipative dynamics of a quantum degenerate gas confined in a one-dimensional harmonic trap, governed by a Lindblad master equation that is quartic in the field operators and conserves the total particle number. While there exist established methods for quadratic (Gaussian) Lindbladians~\cite{prosen2008third,horstmann2013noise,barthel2022solving,thompson2023field}, no general solution strategies are currently available for interacting, quartic Lindbladians. Such generators have nevertheless attracted increasing interest in recent years, including Sachdev--Ye--Kitaev-type models and other interacting open quantum systems~\cite{sa2022lindbladian,kulkarni2022lindbladian,liu2025dissipative,iemini2016dissipative,costa2023spectral,altland2021symmetry,wang2025closed}.

The paper is organized as follows: Section~\ref{sec:sec2} introduces the field-theoretic Lindblad formalism for collective decay of indistinguishable particles in a harmonic trap. We then analyze the resulting dynamics across three regimes: the \emph{tight-trap limit} (Section~\ref{sec:sec3}), where recoil is suppressed and exact solutions are possible; the \emph{Lamb--Dicke regime} (Section~\ref{sec:sec4}), which accounts for leading-order motional transitions and transport; and the \emph{general-recoil regime} (Section~\ref{sec:sec5}), where we evaluate the breakdown of statistical effects at short times.

\section{Field theory for the quantum optics of a degenerate quantum gas}\label{sec:sec2}
In the quantum degenerate regime, the usual quantum optical treatment, describing atoms as distinguishable two level systems with excited and ground states, is insufficient. Once the de Broglie wavelength of each particle is large enough such that motional wavepackets of different particles overlap, a description within second quantization is required. We thus introduce field operators $\hat{\Psi}^\dagger_e(\bm R)$, which creates atoms at $\bm R$ in excited electronic state and $\hat{\Psi}^\dagger_g(\bm R)$, creating atoms of ground type as derived in Ref.~\cite{lewenstein1994quantum}. The bosonic or fermionic character of the particle is encoded in the usual relations
\begin{equation}
	\left[\hat{\Psi}_\alpha(\bm R), \hat{\Psi}_\beta^\dag(\bm R') \right]_\zeta = \delta(\bm R - \bm R')\delta_{\alpha,\beta} \,,
    \label{eq:anti-commutation-free-fields}
\end{equation}
where $\alpha,\beta \in \{e,g\}$, $\zeta = +1$ for bosons and $\zeta = -1$ for fermions, i.e., $[\hat X,\hat Y]_{\zeta} = \hat X \hat Y - \zeta \hat Y\hat X$.

We consider harmonic trapping in all three spatial dimensions, with $\phi_{\bm n}(\bm R)$ denoting the standard harmonic eigenfunctions with quantum numbers $\bm n = (n_x, n_y, n_z)$. We transform the free field operators from the position basis to the trap basis to obtain the operators 
\begin{equation}
  \label{eq:trap_basis_def}
  \begin{aligned}
    \g{\bm n}&=\int\diff{\bm R}\left[\phi_{\bm n}(\bm R)\right]^*\hat{\Psi}_g(\bm R),\\
    \e{\bm n}&=\int\diff{\bm R}\left[\phi_{\bm n}(\bm R)\right]^*\hat{\Psi}_e(\bm R).
  \end{aligned}
\end{equation}
The operators $\g{\bm n}$/$\e{\bm n}$ destroy a particle of the ground/excited-type in the trap level $\bm n$. The resulting (anti-) commutation relations are analogous to the free fields in Eq.~\eqref{eq:anti-commutation-free-fields}, but with the continuous delta-function replaced by a Kronecker delta $\delta_{\bm n, \bm m}$.

Following standard quantum optics techniques ~\cite{Lehmberg70,breuer2002theory,gardiner2004quantum,fazio2025many} we obtain a Lindblad master equation
\begin{equation}
    \dv{\rho}{t}=-\frac{\iu}{\hbar}\commutator{H_\text{at} +H_\text{dd}}{\rho}+ \Lm[\rho].
    \label{eq:master-eq}
\end{equation}
The Hamiltonian $H_\text{at}$ is diagonal in the trap basis. With $\tf_{\bm n}$ the trap frequency for level $\bm n$ and $\omega_0$ the electronic energy in the excited state we obtain
\begin{equation}
      H_\text{at}=  \sum_{\bm n}\hbar\left[\tf_{\bm n}\gd{\bm n}\g{\bm n}+(\omega_0+\tf_{\bm n})\ed{\bm n}\e{\bm n}\right].
      \label{eq:H_at}
\end{equation}
We assume harmonic trapping and the frequencies are \(\tf_{\bm n}=\nu_x n_x+\nu_y n_y+\nu_z n_z\). For the description of light-matter interactions, we introduce the field operator analog to the Pauli ladder operator
\begin{equation}
\Sig{\bm n,\bm n'}=\gd{\bm n}\e{\bm n'}\,, 
\label{eq:ladder-operator-general}
\end{equation}
which annihilates an excited particle in state $\bm{n}'$ and creates a ground particle in state $\bm n$, where the recoil from photon emission changes the motional state from $\bm n'$ to $\bm n$ [see Fig.~\ref{fig:fig1}b)]. From the elimination of the electromagnetic vacuum, we obtain the dipole-dipole Hamiltonian
\begin{equation}
    H_\textrm{dd} = \sum_{\substack{\bm n,\bm n'\\\bm m,\bm m'}}\hbar J_{\bm n,\bm n'}^{\bm m,\bm m'} \Sigd{\bm n,\bm n'}\Sig{\bm m,\bm m'}\,,
    \label{eq:H_dd}
\end{equation}
and the dissipator
\begin{equation}
\begin{aligned}
\Lm[\rho] =\sum_{\substack{\bm n,\bm n\\\bm m,\bm m'}}&\Gamma_{\bm n,\bm n'}^{\bm m,\bm m'}\Big[\Sig{\bm n,\bm n'}\rho\Sigd{\bm m,\bm m'}\\
&-\frac{1}{2}\left(\Sigd{\bm m,\bm m'}\Sig{\bm n,\bm n'}\rho + \rho\Sigd{\bm m,\bm m'}\Sig{\bm n,\bm n'}\right)\Big]\,,
\end{aligned}
	\label{eq:Lindblad general}
\end{equation}
where we define the matrix elements of the coherent and incoherent dipolar interaction in the trap basis in terms of the dipolar Green's function $\mathbf{G}_{\bm n,\bm n'}^{\bm m,\bm m'}$. The dipolar Green's function encodes spatial- and angular dependence of the effective matter interactions and is the microscopic origin of collective radiance. Specifically, the coherent and dissipative matrix elements are related to the dipolar Green's function in the following manner

\begin{subequations}
\begin{align}
	J_{\bm n,\bm n'}^{\bm m,\bm m'} &= -\mu_0 \omega_0^2\Re\left[ \bm d^\dag \mathbf{G}_{\bm n,\bm n'}^{\bm m,\bm m'}\bm d \right]\\
	\Gamma_{\bm n,\bm n'}^{\bm m,\bm m'} &= 2\mu_0 \omega_0^2\Im\left[\bm d^\dag \mathbf{G}_{\bm n,\bm n'}^{\bm m,\bm m'} \bm d \right]\,,
\end{align}
\end{subequations}
where $\mu_0$ is the vacuum permeability and $\bm d$ is the dipole vector of the atoms. The unitary interaction term $J_{\bm n,\bm n'}^{\bm m,\bm m'}$ is a dipolar interaction between two atoms in the system where an excitation is exchanged, thereby changing the motional states of the atoms from $\bm n$ to $\bm n'$ and from $\bm m'$ to $\bm m$ [see Fig.~\ref{fig:fig1}c)]. The dissipative tensor $\Gamma_{\bm n,\bm n'}^{\bm m,\bm m'}$ encodes the same recoil structure, but the excitation is lost from the system instead of being exchanged. We derive exact closed-form expressions for the decay tensor $\Gamma_{n_z,n_z'}^{m_z,m_z'}$ in an effective one-dimensional setting, where the particles are delta-localized along the other two spatial dimensions [see App.~\ref{App:matrix_eta} Eq.~\eqref{eq:Appendix_Gammatensor_general}].

The matrix elements of the dipolar Green's function in the trap basis are defined as
\begin{equation}
\begin{aligned}
	\mathbf{G}_{\bm n,\bm n'}^{\bm m,\bm m'}&=\frac{1}{(2\pi)^3}\int\diff{\bm k}\eta_{\bm n, \bm n'}^*(\bm k)\eta_{\bm m,\bm m'}(\bm k)\widetilde{\mathbf{G}}(\bm k)\, ,
\end{aligned}
\label{eq:GreensFunctionElements}
\end{equation}
with the standard free space dipolar Green's function $\widetilde{\mathbf{G}}(\bm k)$ introduced in App.~\ref{App:Elimination}. We defined the Franck-Condon factors $\eta_{\bm n, \bm m}(\bm k)$ as the overlap between the harmonic oscillator wavefunctions $\phi_{\bm n}$ and $\phi_{\bm m}$ under displacement with momentum $\bm k$
\begin{equation}
\begin{split}
\eta_{\bm n,\bm m}(\bm k)&=\int\diff{\bm R}\me^{-\iu\bm k \bm R}\phi^*_{\bm n}(\bm R)\phi_{\bm m}(\bm R)\,,
\end{split}
    \label{eq:Franck-Condon}
\end{equation}
which encodes the changes in a motional state due to photon recoil. The Franck-Condon factors have a well-known closed form in terms of associated Laguerre Polynomials \cite{cahill1969ordered}, which we give in App.\ref{App:matrix_eta}.\\

\textbf{One-dimensional reduction:} For the sake of simplicity we reduce the quantum numbers for the 3-dimensional harmonic oscillator $\bm n = (n_x, n_y, n_z)$ to a single quantum number by considering only a single relevant trapping direction in $z$ direction by freezing the motion in the two transversal directions $x$ and $y$ and assume the transition dipole moment to also lie in the $z$ direction. We will call $\tf=\tf_z$ for brevity (and equivalently for other quantities such as $k=k_z$) from now on and assume Kronecker delta-conditions on the $x$ and $y$ components of the matrix elements of $\eta_{\bm n,\bm m}$ and other matrix elements.
We structure this paper as an expansion in the Lamb-Dicke parameter $\eta=k_0z_{\text{zpm}}$. Physically, this is an expansion in the localization of the atoms $z_{\text{zpm}}$ compared to the wavelength of the atomic transition $2\pi/k_0$.\\

\textbf{Initial state preparation protocol:} We consider a state preparation scheme utilizing a $\Lambda$-type electronic level configuration consisting of two stable ground manifolds, $\ket{g}$ and $\ket{a}$, and a single common excited state $\ket{e}$. Initially, the system is prepared in a thermal mixture of the two stable ground states. An incoherent, broad-band drive is then applied to the $\ket{a} \rightarrow \ket{e}$ transition. Because the pump is broad-band relative to the trap frequency and the internal excitation process is faster than motional relaxation, the $\ket{e}$ population effectively inherits the thermal distribution of the auxiliary ground state $\ket{a}$. This protocol ensures that inter-species coherences are eliminated, justifying the treatment of the initial state $\rho(t=0;T)$ as a statistical mixture of two independent canonical ensembles for species $\ket{e}$ and $\ket{g}$ at some temperature $T$ [for details see App.~\ref{App:thermal-density-matrices}]. The resulting density matrix $\rho(t=0;T)$ is a mixture of fermionic or bosonic Fock states:
\begin{equation}
\ket{\Ne_0,\Ng_0;\Ne_1,\Ng_1;\ldots;\Ne_n,\Ng_n;\ldots }\,,
\end{equation}
where $\Ne_n$ and $\Ng_n$ denote the number of excited and ground atoms in the trap level $n$, respectively. We define the total excited and ground state numbers as $\Ne=\sum_n \Ne_n$ and $\Ng=\sum_n \Ng_n$, with the total particle number $\Nm=\Ng+\Ne$.\\

\textbf{Intensity as a characteristic signature of superradiance:}
A particular focus in this paper is on the characteristic signature of superradiance, the radiated intensity of the purely dissipative system
\begin{equation}
    I(t) = \hbar \omega_0\sum_{\substack{n,n', m,m'}}\Gamma_{n,n'}^{m,m'} \text{Tr}\left[\Sigd{n,n'}\Sig{m,m'}\rho(t)\right],
\label{eq:intensity_general}
\end{equation}
where $\text{Tr}[\circ]$ denotes the trace. In the standard Dicke treatment, the radiated intensity takes the form of an intense burst of radiation at time $\ln(\Nm)/(\Nm \gamma)$ with a maximum intensity $\propto\Nm^2$~\cite{Dicke54,Gross82}, when considering a fully excited initial state. Also of interest is the intensity at $t=0$, which we refer to as the instantaneous intensity. At finite temperature $T$ it takes the form
\begin{equation}
    I(t=0; T) = \hbar\omega_0\sum_{n,m}\Gamma_{n,m}^{n,m}\Ne_{n}\left(1 + \zeta \Ng_{m}\right)\,.
\label{eq:instantaneous_intensity_general}
\end{equation}
The temperature dependence is implicit in the population of the trap levels. To obtain the initial populations of the trap levels $\Ne_{n},\Ng_{n}$ at finite temperature and with fixed total particle number $\Nm$, we perform Metropolis Monte Carlo sampling of the canonical partition function [for details see App. \ref{App:thermal-density-matrices}]. We will, from now on, drop the global prefactor $\hbar \omega_0$ in the intensity. The linear term in $\Ne_n$ describes independent emission, while the product term $\Ne_n \Ng_m$ encodes statistical modifications of the decay rate, Pauli blocking for fermions and bosonic stimulation for bosons. The instantaneous intensity does not capture the subsequent dynamics at $t>0$, since coherences are generated under time evolution. The intensity at $t=0$ is equivalent to calculations using Fermi's golden rule used in Refs.~\cite{helmerson1990radiative,busch1998inhibition}.

\section{Tight-trap regime}\label{sec:sec3}

In the tight-trap regime, the atomic localization is significantly smaller than the electronic transition wavelength. In this limit, the photon recoil is insufficient to change the motional state of the particles, such that an excited particle in trap level $n$ can only decay into a ground particle in the same level. We thus approximate the Franck-Condon factors in Eq.~\eqref{eq:Franck-Condon} to zeroth order in the Lamb-Dicke parameter $\eta$
\[
\eta_{n,m}(k)=\delta_{n,m} +\mathcal O(\eta).
\]
This essentially removes all geometric information of the particle distribution, placing all emitters onto a single point, resulting in a permutationally symmetric 0-dimensional model. The matrix elements of the interaction thus become $J_{n,n'}^{m,m'} = J_0\delta_{n, n'}\delta_{m, m'}$ and $\Gamma_{n,n'}^{m, m'} = \Gamma_0 \delta_{ n, n'}\delta_{m, m'}$, where $J_0$ is the single particle Lamb-shift and $\Gamma_0=\gamma$ the single particle free space decay rate.

In the tight-trap regime, there is only a single collective collapse operator
\begin{equation}
\Sig{0}=\sum_{n=0}^\infty\gd{n} \e{n}\,.
\label{eq:Sigma0}
\end{equation}
The Hamiltonian and Lindbladian then take the form 
\begin{subequations}
\label{eq:full_tigh_trap_model}
\begin{align}
    \label{eq:Hamiltonian-tigh-trap}
    H &= H_{\rm{at}}+J_0\Sigd{0}\Sig{0}\\
    \Lm[\rho] &=\gamma\left(\Sig{0}\rho\Sigd{0}-\frac{1}{2}\anticommutator{\Sigd{0} \Sig{0}}{\rho}\right)\,,
    \label{eq:Lindblad-tight-trap}
\end{align}
\end{subequations}
where $H_\text{at}$ is the diagonal motional part as defined in Eq.~\eqref{eq:H_at}. The model defined by Eqs.~\eqref{eq:full_tigh_trap_model} now has not only a global U(1) symmetry, but also a local U(1) symmetry, i.e., it is invariant under the transformations
\[
\e{n}\rightarrow \e{n}\me^{-\iu \varphi_n},\;\; \g{n}\rightarrow\g{n}\me^{-\iu \varphi_n}\,.
\]
Thus, not only is the total particle number conserved in the tight-trap regime, but also the particle number on each trap level. When calculating the dynamics of this model, we will utilize the conservation of the local particle number to map the dynamics onto a dissipative collective spin model. For any initial condition, the local particle number conservation allows us to identify a local spin encoded on each trap level, which is conserved.\\
The energy $J_0$ is inconsequential for the dynamics of the system, since it simply splits the collective energy levels by the frequency $J_0$, which does not affect the dynamics so long as the initial state is a thermal density matrix.\\
Considering the two axes of symmetry shown in Fig.~\ref{fig:fig1}\,d)\,e), the tight trap regime corresponds to full permutational symmetry of the Lindbladian, fixing the system along this axis of symmetry. Thus, we are restricted to move along the other axis of symmetry, the symmetry of the exchange of indistinguishable particles. We use temperature as a control parameter to interpolate between the fully quantum-degenerate regime, where exchange symmetry is maximally restrictive, and a dilute regime in which particles effectively behave as distinguishable emitters despite remaining indistinguishable at the operator level.

\begin{figure}
    \centering
    \includegraphics[width=\linewidth]{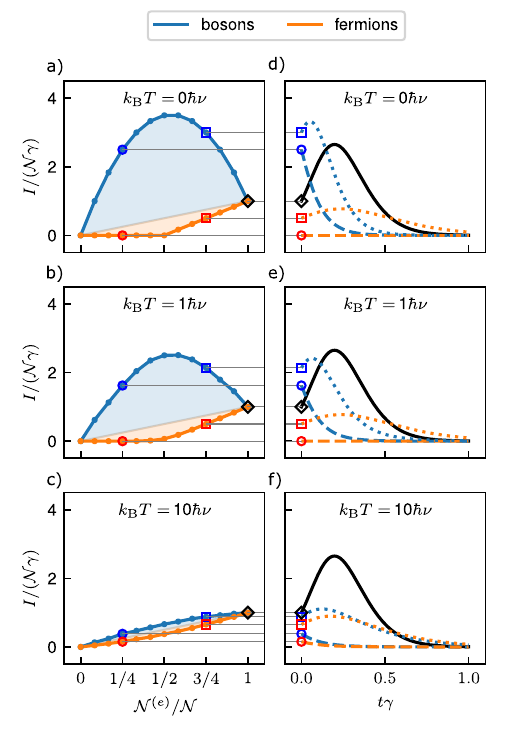}
    \caption{\textbf{Thermal crossover from quantum-statistical to distinguishable radiance:} (a--c) Instantaneous intensity $I(0)$ for $\Nm=12$ (Bosons: blue; Fermions: orange). 
At $T=0$ (a), bosons follow quadratic Dicke scaling while fermions exhibit subradiant suppression, with a larger slope after $\mathcal N^{(e)}/\mathcal N=1/2$. Increasing temperature (b--c) drives the system toward the linear distinguishable limit as motional level co-occupation is diluted. (d--f) Time evolution $I(t)$ for inversions $1/4$ ($\bigcirc$), $3/4$ ($\square$), and $1$ (\rotatebox[origin=c]{45}{$\square$}). A superradiant burst requires $\Ne > \Nm/2$ (initial state above the Bloch sphere equator), occurring here for $\Ne=3\Nm/4$. For full initial inversion ($\Ne=\Nm$), the dynamics are independent of statistics (black line).}
    \label{fig:fig2}
\end{figure}

\subsection{Instantaneous intensity from thermal states}
First, we calculate some observables for thermal states of the quantum gas. This indicates the dynamics of the system at $t=0$, i.e., the initial/instantaneous intensity and correlations. The instantaneous intensity for thermal states in the tight trap regime reduces from the general case in Eq.~\eqref{eq:instantaneous_intensity_general} to
\begin{equation}
    I = \gamma\expval{\Sigd{0}\Sig{0}} =  \gamma\Ne +\gamma\zeta\sum_{n=0}^{\infty}\Ne_n\Ng_n\,,
    \label{eq:Intensity_Bose_Fermi_tight_trap}
\end{equation}
where we dropped the constant prefactor $\hbar \omega_0$. The bosonic and fermionic nature enters both in the particle statistics ($\zeta = \pm 1$) and in the effect of temperature, implicit in the populations $\Ne_n, \Ng_n$.\\

\textbf{Bosons: } Starting from the simplest case at $T=0$, where all bosons are in the ground state of the trap, we obtain the instantaneous intensity
\begin{equation}
    I_\text{B}(T=0)= \gamma \Ne(\Nm - \Ne + 1)\, .
\label{eq:instantaneous_intensity_tight_trap_bosons}
\end{equation}
The quadratic scaling law in Eq.~\eqref{eq:instantaneous_intensity_tight_trap_bosons}, illustrated in Fig.~\ref{fig:fig2}\,a), is the same as in Dicke superradiance. The maximal instantaneous intensity $\propto \Nm^2/4$ is obtained for $\Nm^{(e)}_0 = \Nm^{(g)}_0 = \Nm/2$, which is the classical Dicke result.

When the temperature is increased, particles occupy states beyond the ground state of the trap, and the instantaneous intensity becomes Eq.~\eqref{eq:Intensity_Bose_Fermi_tight_trap} with $\zeta=+1$. Therefore, the intensity decreases monotonically with temperature as $\sum_n \Ne_n \Ng_n \leq \Ne\Ng$. In the limit $T\rightarrow \infty$, the probability of a ground and an excited particle occupying the same level vanishes, reducing the intensity to $I\propto\gamma\Ne $, describing independent particles. We illustrate this in Fig.~\ref{fig:fig2}~c) where the scaling of the instantaneous intensity approaches the linear relation of distinguishable particles at large temperature. The intermediate regime for small temperature $\kb T=\hbar\nu$ is shown in Fig.~\ref{fig:fig2}b), governed by softening of the Bose-Einstein distribution across other levels, reducing the bosonic enhancement with temperature.\\

\textbf{Fermions: } For $T=0$ trap levels are filled under the restriction of the Pauli-exclusion principle. This means that at $T=0$ in the tight-trap regime, no intensity is radiated from fermions below half excitation, as all levels occupied by an excited particle are also occupied by a ground particle, blocking the decay of the excited particle. Therefore, at $T=0$, the instantaneous intensity radiated from fermions in a thermal state is
\begin{equation}
    I_\text{F}(T=0) =
    \begin{cases}
        0 &\Nm^{(e)} \leq \Nm^{(g)}\\
        \gamma(2\Ne - \Nm) &\text{else}
        \,.
    \end{cases}
\end{equation}
For $\Ne > \Ng$ the slope is increased $\propto 2\Ne$ compared to distinguishable particles $\propto \Ne$, which is shown in Fig~\ref{fig:fig2}\,a). This occurs because when the excitation fraction is increased (above half excitation), it simultaneously decreases the number of blocked excitations. 
When the temperature is increased, the probability of a ground and an excited particle occupying the same level decreases monotonically, corresponding to a softening of the Fermi surface [see Fig~\ref{fig:fig2}\,b)]. The intensity is then calculated as Eq.~\eqref{eq:Intensity_Bose_Fermi_tight_trap} with $\zeta=-1$. Thus, for fermions, the intensity increases monotonically with temperature, as opposed to bosons, where it decreases monotonically. For $T\rightarrow \infty$, as for bosons, the probability of more than one particle on a single level vanishes, resulting in the same instantaneous intensity as distinguishable particles $I\propto \gamma\Ne$ [see Fig.~\ref{fig:fig2}\,c)].\\

\textbf{Distinction between dynamical intensity and instantaneous intensity: } It is important to distinguish the instantaneous intensity at some temperature as a function of the excitation fraction from the dynamical intensity, starting from some initial excitation fraction at some temperature. To highlight this point, we show in Fig.~\ref{fig:fig2}d)-f) the dynamical intensity starting from a few selected thermal initial states at different inversions. It shows that while initial correlations are reduced at finite temperature, correlations can still be built up dynamically due to the permutationally symmetric dissipator. We go into more details of the dynamical behavior in the next subsection.

\subsection{Dissipative dynamics --- spin mapping}

\begin{figure*}[!ht]
    \centering
    \includegraphics[width=\linewidth]{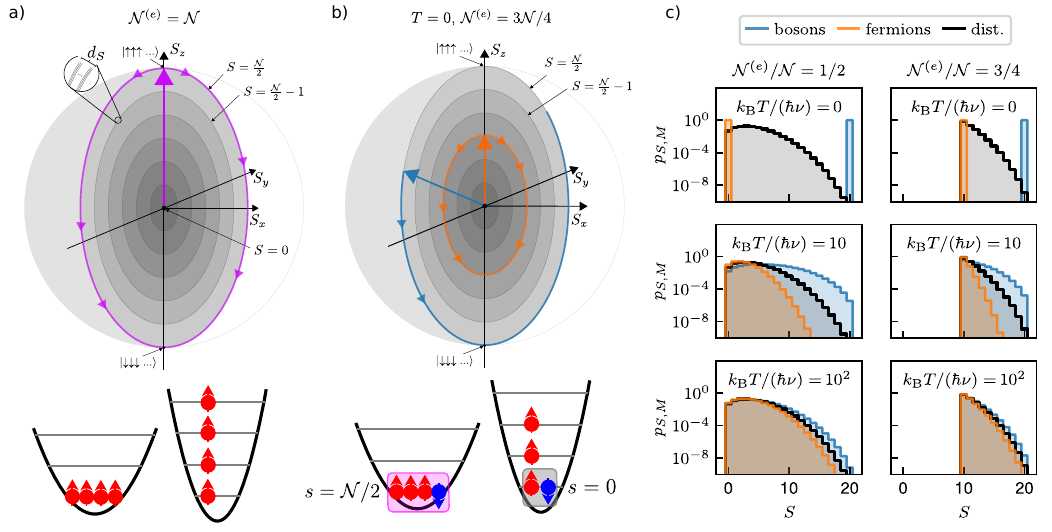}
    \caption{\textbf{Mapping quantum statistics to collective spin shells.} a) At full inversion ($\Ne=\Nm$), all particles occupy a unique non-degenerate state; dynamics are restricted to the fully symmetric Dicke subspace ($S=\Nm/2$) regardless of statistics or temperature. b) At $T=0$, bosons (blue) remain in the $S=\Nm/2$ shell. Fermions (orange) occupy a reduced spin shell $S=|\mathcal{N}^{(e)} - \mathcal{N}^{(g)}|/2$ because fully filled lower levels form spin-singlets. c) At finite temperature, a thermal state contains a mixture of population across multiple spin shells. In the high-temperature limit ($k_\textrm{B} T \gg \hbar\nu$), the spin-shell distributions for bosons and fermions converge as exchange symmetry becomes irrelevant.}
    \label{fig:fig3}
\end{figure*}

Both the dissipative dynamics and the instantaneous intensity can be intuitively understood via a spin mapping procedure. The tight-trap regime admits a complete analytical description through a mapping of the bosonic and fermionic algebra onto the usual SU(2) spin algebra, with details in App.~\ref{App:spin_decomposition}. Because the collective operator $\Sig{0}$ acts identically on each trap level, the dissipator is invariant under trap-level permutations. Furthermore, the local particle number $\gd{n}\g{n} + \ed{n}\e{n}$ is conserved on each level $n$. This allows us to define local spin operators:
\begin{align*}
    \hat{s}^{(x)}_n&=(\ed{n}\g{n}+\gd{n}\e{n})/2\,,\\
    \hat{s}^{(y)}_n&=(\ed{n}\g{n}-\gd{n}\e{n})/(2\iu)\,,\\
    \hat{s}^{(z)}_n&=(\ed{n}\e{n} - \gd{n}\g{n})/2\,,
\end{align*}
where the spin magnitude $s_n$ is conserved and fixed by the local occupation and particle statistics
\begin{equation}
        s_{n} = 
        \begin{cases} 
        \left(\Nm^{(e)}_n + \Nm^{(g)}_n\right) /2 &\text{bosons} \\[2ex]
        \left| \Nm^{(e)}_n - \Nm^{(g)}_n \right| /2 &\text{fermions.}
        \end{cases}
\end{equation}
The total Hilbert space $\Hm_\text{tot}$ decomposes into collective spin sectors $\mathcal{H}_{\text{tot}} \cong \bigoplus_{S, \alpha} \mathcal{H}^{(\alpha)}_S$, where $\alpha$ labels the multiplicity $d_S$ of the subspace with spin $S$. Thus, each collective spin state is defined via the magnitude of the collective spins $S$, the magnetization/inversion $M$, and the multiplicity number $\alpha$. Since the collective operators $\hat{\bm S}_0 = \sum_n \hat{\bm s}_n$, with $\hat{\bm s}_n=(\hat{s}^{(x)}_n,\hat{s}^{(y)}_n,\hat{s}^{(z)}_n)^T$, act diagonally on the index $\alpha$, we compute the intensity by projecting the initial density matrix onto the collective spin populations $p_{S,M} = \sum_{\alpha} \langle S,M,\alpha | \rho | S,M,\alpha \rangle$. Then, the radiated intensity as a function of time is given by
\begin{equation}
I(t) = \gamma \sum_{S,M} p_{S,M}(t) (S+M)(S-M+1)\,.
\end{equation}
The time evolution of each population $p_{S,M}$ follows the standard Dicke master equation (which can be solved analytically~\cite{lee1977exact1,lee1977exact2,holzinger2025compact, holzinger2025solving})
\begin{equation}\label{eq:dicke_short}
\dot p_{S,M} = -\gamma h_{S,M} p_{S,M} + \gamma h_{S,M+1} p_{S,M+1}\,,
\end{equation}
with $h_{S,M} = (S+M)(S-M+1)$. This mapping reduces the dimension of the Hilbert space from exponential to at most quadratic in the particle number $\Nm$. As illustrated in Fig.~\ref{fig:fig3}c), the influence of temperature and statistics is entirely encoded in the initial distribution $p_{S,M}(0)$. Bosons favor larger total spins (maximal at $T=0$), while fermions form spin-0 singlets on doubly occupied levels (one ground and one excited particle), suppressing collective decay.

We illustrate this decomposition for two paradigmatic cases in Fig.~\ref{fig:fig3}a,b). In Fig.~\ref{fig:fig3}a) we consider the fully excited state, so that the decomposition into the collective spin components only obtains a contribution from $S=\Nm/2$, the non-degenerate and completely symmetric sector ($d_{\Nm/2}=1$). In Fig.~\ref{fig:fig3}b), we consider $T=0$, but now where not all particles are initially excited. Then, bosons still map onto a spin $\Nm/2$, but fermions map onto a smaller spin, as states occupied by both a ground and an excited particle form a spin of length 0 (singlet). At non-zero temperature the spin decomposition results in a distribution over many spin sectors [see Fig.~\ref{fig:fig3}c)]. Much of the complexity of the initial state is now hidden in the degeneracy within the spin manifold, which is inconsequential for the time evolution since only the weight of the projection onto this spin sector matters for the calculation of collective observables. 

In the high temperature limit, where there is only ever one particle per trap level, the decomposition can be obtained from the addition of only two spins. For $(\Ne - \Ng)/2 = M$, the initial populations are given by
\begin{equation}
\label{eq:CGs_infty}
p_{S,M}= \left|\bra{S,M}\left[\ket{\tfrac{\Ne}{2},\tfrac{\Ne}{2}}\otimes\ket{\tfrac{\Ng}{2},-\tfrac{\Ng}{2}}\right]\right|^2\,,
\end{equation}
and can therefore be obtained in closed form from the Clebsch-Gordan coefficients, which give a non-zero contribution to all collective spin states $S$ for which $|M|\leq S \leq \Nm/2$. We will see that much of the physics can be understood by the length of the collective spin, which for distinguishable particles is given by
\begin{equation}
    \langle \hat{\bm{S}}_0^2\rangle = \sum_{S=|M|}^{\Nm/2}S(S+1) p_{S,M} = M^2 + \frac{\Nm}{2}\,. 
\label{eq:S^2_dist}
\end{equation}
Scenarios of this type, where an ensemble of emitters is split into two sub-ensembles, have recently been investigated in the context of sensing and energy transfer~\cite{kaubruegger2025lieb, padilla2025generating, fasser2024subradiance}.

\begin{figure*}[!t]
    \centering
    \includegraphics[width = \textwidth]{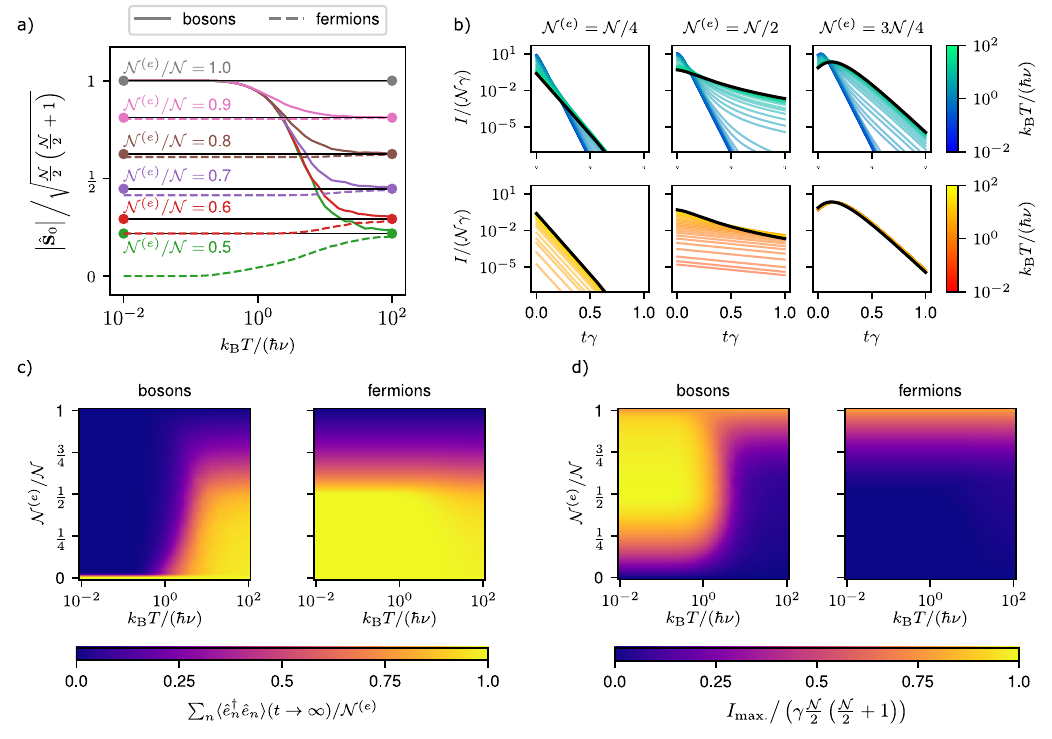}
    \caption{\textbf{Thermal crossover of collective spin and radiance.}  a) Collective spin length as a function of temperature. Statistical effects vanish at $\kb T \gg \hbar\nu$, where both species converge to the distinguishable limit (horizontal lines) obtained from Eqs.~\eqref{eq:CGs_infty},\eqref{eq:S^2_dist}. Due to symmetry of the Clebsch-Gordan coefficients, the collective spin decomposition is invariant under $\mathcal{N}^{(e)} \leftrightarrow \mathcal{N}^{(g)}$.
b) Time-dependent intensity $I(t)$ for varying inversions and temperatures for bosons (top) and fermions (bottom). Dynamics converge to the distinguishable limit (black) from above (bosons) or below (fermions).
c)--d) Residual excitations and peak intensity as a function of $T$ and $\mathcal{N}^{(e)}$. At high temperatures, observables become species-independent as level co-occupation is suppressed. Calculations use $10^5$ thermal state samples per point and $\Nm = 40$.}
    \label{fig:fig4}
\end{figure*}

\textbf{Bosons: } For bosons at $T=0$ the initial state lies in the fully permutationally symmetric manifold 
\begin{equation}
    \begin{split}
        S&=(\Nm^{(e)}_0 + \Nm^{(g)}_0)/2 =\Nm/2\\
        M&= (\Nm^{(e)}_0 - \Nm^{(g)}_0)/2\,.
    \end{split}
\end{equation}
Thus, the dynamics are constrained to the fully symmetric manifold where the number of excited particles fixes the initial state within this subspace.
Most of the physics of the nonzero temperature case can be understood from studying the spin projection of thermal density matrices and observing the dependence of the $S$-distribution on temperature and inversion, as illustrated in Fig.~\ref{fig:fig3}c). The fully inverted ensemble is always completely symmetric $S=\Nm/2$, and therefore independent of temperature. Only when $T>0$ and not all particles are initially excited $\Ne < \Nm$, do we obtain a decomposition over spin shells. The decomposition is symmetric with respect to reflection about the equator of the Bloch sphere, such that, for example, $\mathcal N^{(e)}/\mathcal{N}=1/4$ and $\mathcal N^{(e)}/\mathcal{N}=3/4$ are identical. Notably, the fermionic and bosonic distributions are identical as $T\to\infty$, indicating the distinguishable particle limit. At $T=0$ both fermions and bosons map onto a single collective spin value, while the distribution for distinguishable particles is broad (given the restriction $S\geq|M|$). The fermionic distribution broadens monotonically, while the bosonic distribution first broadens and then decreases again.\\
The dynamics can be mostly understood by calculating the mean spin length $| \hat{\bm{S}}_0 |= \sum_{S=|M|}^{\Nm/2} \sqrt{S(S+1)}p_{S,M}$ as a function of temperature, and for the different initial excitation fractions. We show in Fig.~\ref{fig:fig4}\,a) how the bosonic spin expectation value decreases from $S=\Nm/2$ to the distinguishable case at large temperatures. Increasing the temperature distributes the particles over many trap levels, breaking up the large spin into several smaller spins. It is therefore clear that at finite temperature, the radiated intensity is reduced compared to the $T=0$ case, but still exceeds the intensity radiated by distinguishable particles as displayed in Fig.~\ref{fig:fig4}\,b). Further, we compute the number of remaining excitations in the steady-state and the peak intensity for all initial excitation fractions and over a broad range of temperatures in Fig.~\ref{fig:fig4}\,c)\,d). As long as the thermal energy $\kb T$ is small compared to the trap level spacing $\hbar\nu$, the dynamics are mostly superradiant, resulting in the decay of almost all initially excited particles and an enhanced radiated intensity. When the thermal energy and the trap level spacing become comparable, the peak intensity reduces, and some fraction of the initial excitations remains in the steady-state. Once the thermal energy becomes much larger than the trap level spacing, the particles are effectively distinguishable, as illustrated by the observables in Fig.~\ref{fig:fig4}.

\textbf{Fermions: } At $T=0$ fermions fill the trap levels obeying the Fermi exclusion principle. Therefore, in the tight-trap regime, there are no dynamics below half excitation, as each level that is populated by an excited particle is also populated by a ground particle blocking its decay. When the excitation fraction $\Ne/\Nm$ is increased above $1/2$, there exist excited particles that singly occupy trap levels. Thus, each trap level occupied by a single excited particle encodes a spin$-1/2$ degree of freedom, and each doubly occupied level encodes a spin$-0$. Therefore, the populations for $\Ne \geq \Ng$ are
\[
    p_{S,M} = \left|\bra{S,M}\left[\ket{0,0}^{\otimes \Ng} \otimes \ket{\tfrac{1}{2},\tfrac{1}{2}}^{\otimes(\Ne - \Ng)}\right]\right|^2.
\]
Performing the transformation into the collective basis, we obtain the quantum numbers for the collective spin 
\begin{equation}
    \begin{split}
    S &= |\Nm^{(e)} -\Nm^{(g)}|/2\,, \\
    M &= (\Nm^{(e)} -\Nm^{(g)})/2\,,
    \end{split}
\end{equation}
which also holds for $\Ne < \Ng$. The degeneracy quantum number $\alpha$ is not relevant in the case of fermions, as there is no degeneracy. Rather, the collective spin is fixed to a single irreducible representation $\alpha$.
Below half excitation $\Nm^{(e)} \leq \Nm^{(g)}$, the collective spin is already in the lowest state $\ket{S,-S}$, and there are no dynamics $\Sig{0} \ket{S,-S}=0$. The size of the collective spin is reduced compared to the bosonic case, as all doubly occupied states with $\Nm^{(e)}_n = \Nm^{(g)}_n = 1$ have $s_{n} = 0$ and therefore do not contribute to the size of the collective spin [see Fig.~\ref{fig:fig3} b),c)]. This also means that of the initial $\Ne$ excitations, only $\Ne - \Ng$ can decay, and the remaining $\Ng$ excitations will remain in the steady state. We can obtain the maximum intensity and the time of maximum intensity of the superradiant burst, under the usual semiclassical approximations, as $I_\text{max} \propto \gamma S(S+1)$ and $t_\text{max} = \ln(2S)/(2S\gamma)$ \cite{Dicke54,Gross82}.

At finite temperature, we again consider the decomposition of the initial thermal state into the different collective spin components. For fermions, this only ever involves spin$-1/2$ or spin$-0$. Therefore, the collective spin state for fermions at $T>0$ with $n_\uparrow$ trap levels occupied by a single excited particle, $n_\downarrow$ trap levels occupied by a single ground particle, and $n_0$ trap levels occupied by both a ground and an excited particle is given by 
\begin{equation}
    p_{S,M} = \left|\bra{S,M} \left[\ket{\tfrac{1}{2},\tfrac{1}{2}}^{\otimes n_{\uparrow}}\otimes \ket{\tfrac{1}{2},-\tfrac{1}{2}}^{\otimes n_{\downarrow}}\otimes\ket{0,0}^{\otimes {n_0}}\right]\right|^2\,,
    \label{eq:fermi_spin_collective_general}
\end{equation}
with $n_\downarrow + n_0 = \Ng$, $n_\uparrow + n_0 = \Ne$ and $n_{\uparrow} + n_{\downarrow} + 2n_{0} = \Nm$. 
Increasing the temperature reduces the number of blocked excitations (spin singlets) $n_0$, resulting in an increase in the length of the collective spin. However, as long as the temperature remains finite, and $\Ne < \Nm$, there will be a finite probability that trap levels are occupied by a spin--$0$. Thus, the collective spin at finite temperature will always be reduced compared to distinguishable particles. Only when the singlet number is zero $n_0=0$ does the fermionic collective spin given by Eq.~\eqref{eq:fermi_spin_collective_general}, reduce to the expression of distinguishable particles in Eq.~\eqref{eq:CGs_infty}. Therefore, the peak intensity of the superradiant emission will be reduced and delayed (provided $\Ne > \Ng$ which is required for the occurance of a burst). Compared to the bosonic distribution, the fermionic nature thus pulls the distribution to lower $S$ values at finite temperature, hence narrowing the distribution compared to distinguishable particles, as illustrated in Fig~\ref{fig:fig3}c). However, the relative change in the collective spin length as a function of the temperature for fermions is rather small, as illustrated in Fig.~\ref{fig:fig4}a). Therefore, the change in observables, such as peak intensity and remaining excitations, is also rather small [see Fig.~\ref{fig:fig4}c),d)].

In this tight-trap limit, the radiative dynamics are thus a direct reflection of the initial state's spin-shell distribution, where the competition between exchange statistics and thermal dilution determines the weight of the available superradiant decay channels.

\section{The Lamb-Dicke regime}\label{sec:sec4}
In this section, we relax the tight-trap condition and hence the permutational invariance of the Lindbladian and investigate the Lamb–Dicke regime defined by the expansion of the Franck-Condon factors to $\mathcal{O}(\eta^2)$, in which photon recoil during spontaneous emission can induce transitions between neighboring trap levels.

\textbf{Neglecting unitary interactions:} While the unitary dipolar interactions did not affect the dynamics in the tight-trap regime, they can in the Lamb-Dicke regime. However, to isolate the fundamental impact of particle statistics on collective decay rates, we focus on a minimal model of purely dissipative dynamics. Neglecting the unitary $J_{n,n'}^{m,m'}$-tensor allows for a clear mapping of exchange symmetry effects on radiance, unclouded by the effects of the unitary dipolar interactions. In our framework, the non-Hermitian Hamiltonian $H_{\text{nh}}$ already accounts for the breakdown of local symmetries and the resulting transport. The primary qualitative effect of $J$ would be to potentially accelerate the delocalization of excitations across the trap levels, effectively decoupling the transport timescale from the radiative lifetime $1/\gamma$.

In addition to the collective operator $\Sig{0}$, we define the recoil-induced collective operators $\Sig{\pm}$, and the operator $\Sig{\text{lin.}}$
\begin{equation}
\begin{aligned}
    \Sig{-}&=\sum_{n=1}^\infty  \sqrt{n}\gd{n-1}\e{n},\\ \Sig{+}&=\sum_{n=0}^\infty \sqrt{n+1}\gd{n+1}\e{n},\\
    \Sig{\text{lin}}&=\sum_{n=0}^\infty n\gd{n}\e{n}\,,
\end{aligned}
\label{eq:Collective_Operators_Lamb_Dicke}
\end{equation}
with cross terms in the Lindblad superoperator. To identify the independent decay channels, we diagonalize the decay channels to leading order $\mathcal{O}(\eta^2)$ [for more details see App.~\ref{App:LambDicke}]. This yields the following diagonal collective collapse operators
\begin{equation}
\begin{aligned}
\hat{L}_1 &= \sqrt{\gamma}\sum_{n=0}^{\infty}\sqrt{1-\frac{2}{5}\eta^2(2n + 1)}\gd{n}\e{n}\,,\\
\hat{L}_2 &= \sqrt{\gamma\frac{2}{5}\eta^2}\left(\Sig{+} + \Sig{-}\right)\,.
\end{aligned}
\end{equation}
The operator $\hat{L}_1$ describes the primary electronic decay, where the rate is now level-dependent and reduced by the Lamb-Dicke parameter. The operator $\hat{L}_2$ represents transitions that redistribute population between adjacent trap levels. A third operator, $\hat{L}_3 \propto (\Sig{+} - \Sig{-})$, is dark at this order. Note that we perform the perturbative expansion in a way such that the total decay rate for a single excited particle in any trap level sums to $\gamma$. Also note that the definition of the collapse operator $\hat{L}_1$ implies a radius of convergence for the perturbative expansion $\eta^2(2n + 1)<5/2$ for all occupied levels $n$. As such, the expansion is not valid for an arbitrary number of occupied levels, and $\eta$ must be chosen judiciously to remain in a regime where the expansion is physical. Consequently, we restrict the Lamb-Dicke analysis to $T=0$ to ensure expansion validity and restrict the emitter number for fermionic systems. This limit isolates the breakdown of permutational symmetry while maximizing the dynamical impact of particle statistics, focusing the investigation on a single axis of symmetry competition [see~Fig.~\ref{fig:fig1}d,e)].
\begin{figure*}[!t]
    \centering
    \includegraphics[width=\linewidth]{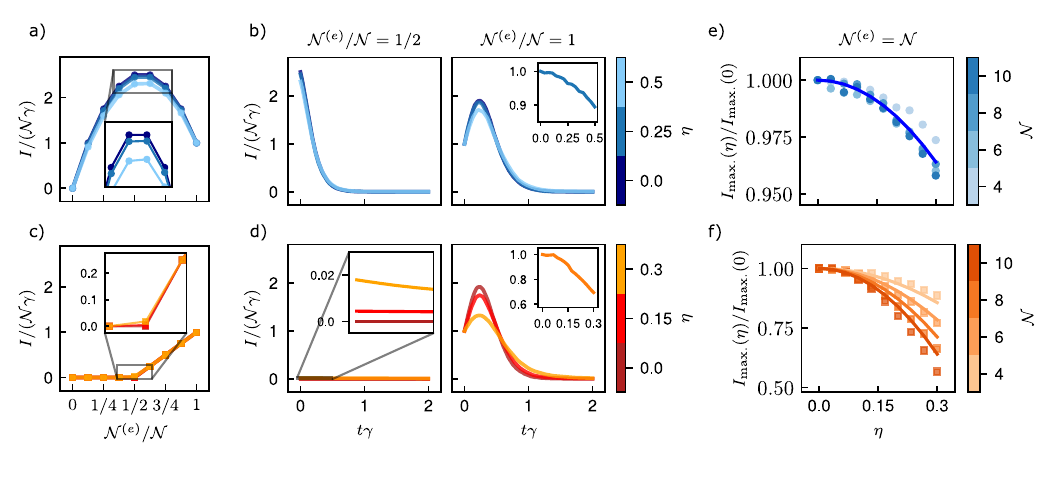}
    \caption{\textbf{Weakening of collective radiance in the Lamb-Dicke regime.} a),b) Bosons at $T=0$: Increasing the trap width $\eta$ monotonically reduces the instantaneous intensity [Eq.~\eqref{eq:inst_int_bose_lamb_dicke}] and suppresses the superradiant burst. c),d) Fermions at $T=0$: The instantaneous intensity remains largely invariant with $\eta$, except near half-excitation (see inset), confirming Eq.~\eqref{eq:int_inst_fermi_lamb_dicke}. Below half-excitation ($\mathcal{N}^{(e)} < \mathcal{N}/2$), dynamics are frozen; at $\mathcal{N}^{(e)} = \mathcal{N}/2$, dynamics are highly subradiant but follow the analytical small-$\eta$ limit. Above half-excitation, the superradiant burst intensity diminishes with $\eta$. Inset d): Peak intensity vs.\ $\eta$ for full inversion, normalized to the $\eta=0$ limit. The numerical simulations in c),d) are performed for $\Nm = 8$ particles and include processes up to $\eta^6$. In e,f) we show how the scaling of the superradiant burst as a function of $\eta$ and for different particle numbers. While for bosons the peak intensity reduces as $\eta^2$, for fermions it reduces as $\Nm\eta^2$. The solid lines are the mean-field result from the main text. To isolate leading order effects, the simulations in e),f) only include processes up to $\eta^2$.}
    \label{fig:fig5}
\end{figure*}

\subsection{Instantaneous intensity from thermal states}
The instantaneous intensity from thermal initial states in the Lamb-Dicke regime takes the general form
\begin{equation}
\begin{split}
        I =\sum_{n=0}^{\infty} \Ne_n&\Big[\left(\Gamma^{(\text{LD})}_0 - 2n\Gamma^{(\text{LD})}_1\right)\left(1 + \zeta \Ng_n\right)\\
    &+ (n+1)\Gamma^{(\text{LD})}_{1}\left(1 + \zeta \Ng_{n+1}\right) \\
    &+ n\Gamma^{(\text{LD})}_{1}\left(1 + \zeta \Ng_{n-1}\right) \Big]\,,
    \label{eq:instantaneous_lamb_dicke_general}
\end{split}
\end{equation}
using shorthands $\Gamma_0^{(\text{LD})}=\gamma(1 - \frac{2\eta^2}{5})$ and $\Gamma_1^{(\text{LD})}=\gamma\frac{2}{5}\eta^2$. This captures how, depending on the population of neighboring trap levels, motional transitions affect the instantaneous intensity.

\textbf{Bosons: }At $T=0$, all particles are initially in the ground state of the trap. Hence, the instantaneous intensity is given by
\begin{equation}
    \label{eq:inst_int_bose_lamb_dicke}
    I_\text{B}(T=0) =\gamma\Ne_0 \left[1 + \left(1 - \frac{2\eta^2}{5}\right)\Ng_0\right]\,.
\end{equation}
Thus, due to presence of an additional decay channel, the instantaneous intensity is reduced compared to the tight-trap regime [see Eq.~\eqref{eq:instantaneous_intensity_tight_trap_bosons}]. This occurs because the on-level decay rate, which is Bose-enhanced at $T=0$, is reduced at the expense of the decay rate into the second trap level, which is not Bose-enhanced. When the total number of ground particles and the total decay rate are fixed, this leads to a monotonous decrease in the instantaneous intensity when increasing $\eta$ [see Fig.~\ref{fig:fig5}\,a)].

\textbf{Fermions: }For fermions below half excitation $\Ne < \Ng$, there is no change in the instantaneous intensity since the decay of all excited particles remains blocked by a ground particle on the same level, the level below, and the level above. At exactly half excitation $\Ne = \Ng$, there is a single excited particle on the trap level $n=\Nm/2-1$, which is no longer blocked from above, i.e., the next highest level $n+1=\Nm/2$ is not occupied by a ground particle. Hence, this excitation decays at the rate $\Gamma_1^\text{(LD)}\Nm/2 = \gamma\eta^2\Nm/5$. Once $\Ne>\Ng$, another effect appears, where the first excited particle, which solely occupies a trap level, has a reduced total decay rate, as it can not decay into the trap level below. This reduction in intensity cancels the gain from the previously blocked particle, which can decay into the trap level above. Thus, at $T=0$ we can summarize the intensity for fermions in the Lamb-Dicke regime as [see Fig.~\ref{fig:fig5}c)]
\begin{equation}
    I_\text{F}(T=0) = 
    \begin{cases}
    0 & \text{if\; $\Ne < \Ng$}\\
     \gamma\eta^2\Nm/5 & \text{if $\Ne = \Ng$}\\
    \gamma(2\Ne - \Nm) & \text{else.}
    \end{cases}
    \label{eq:int_inst_fermi_lamb_dicke}
\end{equation}
Thus, for fermions at $T=0$ , the instantaneous intensity in the Lamb-Dicke regime only differs from the tight-trap regime at exactly half excitation $\Ne = \Ng$.
\subsection{Dynamical super and subradiance and long-range recoil-induced transport}
\begin{figure*}[!t]
    \centering
    \includegraphics[width = \linewidth]{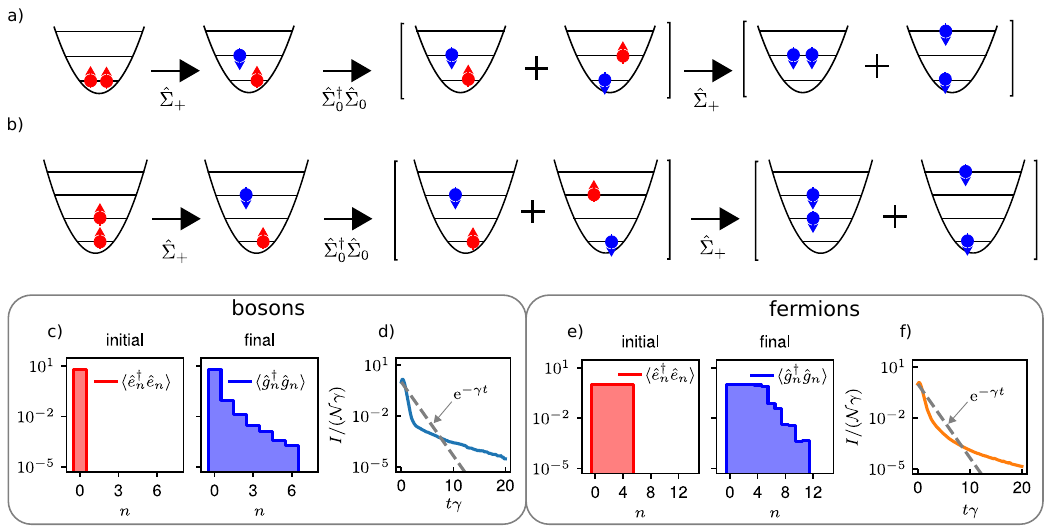}
    \caption{\textbf{Long-range transport and subradiant tails.} a) Illustration of how bosons at $T=0$ are transported by more than one trap level in the Lamb-Dicke regime. For simplicity, we show only part of the resulting state when acting with $\Sig{+}$ and neglect the coefficients. b) Same as in a), but for fermions. c) An initially fully excited ensemble of bosons at $T=0$ shows a long-time distribution with an exponential tail, where particles are distributed over many trap levels. The long-range transport is accompanied by a long subradiant tail in the intensity. d) Same as c), but for fermions. The simulation in c)-f) is performed for $\Nm = 6$ and includes processes up to $\eta^{12}$.}
    \label{fig:fig6}
\end{figure*}

The collective operators $\Sig{\pm}$, which move particles between levels, make a general analytical and numerical description of the dynamics intractable. To understand why the complexity of the problem increases significantly in the Lamb-Dicke regime, we can consider attempting a description in terms of a collective spin as in the tight-trap regime. Since the photon recoil can now move particles between trap levels, the spin encoded on each trap level is no longer conserved. Thus, the collective spin is also no longer conserved. In leading order $\eta^2$, this would result in the coupling of adjacent collective spin shells. However, if the initial state contains particles occupying multiple trap levels (e.g. in the case of fermions or non-zero temperature), the decay rates are level-dependent, thus spoiling a collective spin description where the information of level occupations is removed.

Thus, we consider the time evolution perturbatively in the Lamb-Dicke parameter $\eta$. We hence design an algorithm that constructs a perturbative basis to significantly decrease the dimension of the accessible Hilbert space [for details see App.~\ref{app:perturbative_basis}], which we call symmetry-guided perturbative basis expansion.

\textbf{Bosons: }As discussed earlier, we now perform the time evolution up to a given order in $\eta$, relying on a perturbative basis generation to reduce the relevant Hilbert space dimension. In Fig.~\ref{fig:fig5}b) we show how increasing the Lamb-Dicke parameter $\eta$ modifies the intensity. As $\eta$ is increased, the peak intensity always decreases. Essentially, increasing $\eta$ increases the probability flux out of the symmetric sector, decreasing the intensity of the superradiant burst. This aligns well with the peak intensity shown in Fig.~\ref{fig:fig5}c). To leading order $\propto\eta^2$, one finds for the mean field $\Ng_0=\frac{\Gamma_1^\text{(LD)}}{\gamma}\Ng_1$ so that the peak intensity reduces as $I_{\text{max}}(\eta)\approx (1-\frac{2\eta^2}{5})I_{\text{max}}(0)$ which fits well with the numerical simulation in Fig.~\ref{fig:fig5}e), confirming an $\Nm$-independent leakage channel, reducing the collective spin, via decay into the second trap level.

\textbf{Fermions: } For fermions at $T=0$, below half excitation, there is no change in the dynamics, as the levels below and above any excited particle are occupied. At exactly half excitation, the excited particle in the highest occupied trap state ($n = \Nm/2-1$) can now decay into the next highest level ($\Nm/2-1 \rightarrow \Nm/2$). Considering only processes to order $\eta^2$, the dynamics can then be calculated analytically and are described by a spin-1/2 degree of freedom at the Fermi edge, with an effective decay rate $\Gamma_1\Nm/2$ so that $I(t) = \tfrac{\Gamma_1^\text{(LD)}\Nm}{2} \me^{-\frac{\Gamma_1^\text{(LD)}\Nm}{2} t}$. This has been proposed as a method for qubit encoding \cite{dowdall2017fast}.\\

As for bosons, we perform exact numerical simulations for small system sizes at $T=0$ using a symmetry-guided perturbative basis expansion. In the Lamb-Dicke regime below half excitation at $T=0$ there are no dynamics. At exactly half excitation, the intensity in the Lamb-Dicke regime is non-zero, and therefore larger than in the tight-trap regime, where it is zero. The early dynamics are well described by the analytical approximation keeping only terms up to $\eta^2$, resulting in a simple exponential decay [see Fig.~\ref{fig:fig5}d)]. Above half excitation, the intensity of the superradiant bursts is now reduced. This is most noticeable in the case of initial full excitation, because there the permutationally symmetric component is the smallest. In this case, the Pauli-exclusion principle enforces the occupation of the first $\Nm$ levels, and due to higher trap levels having a larger spatial extent, the permutationally symmetric component is smaller compared to bosons. Thus, despite rather small values of $\eta$, the burst intensity is more strongly reduced for fermions, which is well approximated by the scaling $I_{\text{max}}(\eta)\approx (1-\frac{2\Nm\eta^2}{5})I_{\text{max}}(0)$. Following the same logic as in the bosonic case, this can be derived as the rate over all decay processes that break the permutational symmetry, i.e., move particles between trap levels, relative to the total decay rate.\\

\textbf{Subradiant tails and transport:} One surprising aspect of the Lamb-Dicke regime is the long-range transport of particles in the trap. This arises because any local excitation gets delocalized over all occupied trap levels via the action of the permutationally symmetric non-Hermitian Hamiltonian $\Sigd{0}\Sig{0}$. This can be illustrated using the same example as given by Dicke \cite{Dicke54}, where the two-particle product state with one excitation is expanded in the collective basis. The key difference is that here the indices do not refer to particles, but rather to a trap levels
\begin{equation}
\ket{\uparrow_{n},\downarrow_{n+1}} = \tfrac{1}{\sqrt{2}}\left(\ket{+} + \ket{-}\right)\,,
\end{equation}
with $\ket{\pm} = (\ket{\uparrow_{n},\downarrow_{n+1}} \pm \ket{\downarrow_{n},\uparrow_{n+1}})/\sqrt{2}$. Generalizing this principle, where one repeatedly acts with $\Sig{+}$ followed by a transformation back into the symmetric basis, effectively partially re-excites all ground particles, giving rise to long-range transport of particles in the trap. We illustrate this in Fig.~\ref{fig:fig6}, where we show the transport effect schematically for bosons in Fig.~\ref{fig:fig6}a) and for fermions in Fig.~\ref{fig:fig6}b). Further, in Fig.~\ref{fig:fig6}c),e) we show the long-time distribution of excited and ground particles in the trap, starting from a fully excited initial state at $T=0$, for bosons and fermions. The long-time distribution shows an exponential tail, which is truncated, as our numerical simulation only includes processes up to $\eta^{12}$. The long-range transport is accompanied by a subradiant tail of the intensity, where the symmetric fraction has decayed, and excitations can decay only via repeated application of $\eta^2$ processes, which move particles between trap levels [see Fig.\ref{fig:fig6}\,d),\,f)].

\section{Beyond the Lamb-Dicke regime}\label{sec:sec5}

We now lift the Lamb-Dicke restriction, allowing for recoil-induced transitions between trap levels of arbitrary separation during photon emission. Unfortunately, this makes any numerical simulations of the dynamics impossible, as the Hilbert space does not truncate to a finite dimension. However, using the general, closed-form expressions for the dissipative matrix elements, we can examine the instantaneous intensity and its derivative. From this, we gain insight into the behavior at an arbitrary trap width, in particular, how the system behaves in the thermodynamic limit, when both trap width and particle number are taken to infinity. Indeed, the combination of $I(0)$ and $\dot I(0)$ act as a predictor for the occurrence of superradiant burst because $I(0)$ indicates the existence of superradiant correlations in the initial state and $\dot I(0)$ the buildup of them under the evolution of the Lindbladian. There is no process in the system that would delay the buildup and hence the information at time $t=0$ is well suited to predict the existence and formation of a superradiant burst.

\subsection{Instantaneous intensity from thermal states}
\textbf{Bosons: } The instantaneous intensity now takes the general form stated in Eq.~\eqref{eq:instantaneous_intensity_general}. For the case at $T=0$, this reduces to
\begin{equation}
I_{\text{B}}(T=0)= \Ne_0\left(\gamma + \gamma_\textrm{B}(\eta) \Ng_0 \right)\,,
\end{equation}
with $\gamma_\textrm{B}(\eta)$ the Bose-enhanced rate, which takes the form
\begin{equation}
\gamma_\textrm{B} = \gamma\frac{3\sqrt{\pi}(\eta^2 + 1/2)\erf(\eta) - 3\eta\me^{-\eta^2}}{8\eta^3}\,.
\label{eq:beyond_gamma_B}
\end{equation}
For $\eta=0$ this reduces to the result in the tight-trap regime $I_\textrm{B} = \gamma\Ne_0(1 + \Ng_0)$ and for $\eta\rightarrow\infty$, the Bose-enhanced rate decays as $\eta^{-1}$ and the instantaneous intensity reduces to that of independent emitters $\gamma\Ne_0$ (for $\Ne_0,\Ng_0$ finite). This is expected, as increasing $\eta$ corresponds to a broader trap and therefore a lower density. In the limit $\eta\rightarrow0$ (Dicke limit), the effect of particle statistics is strongest, as all particles are localized at a single point. To see if superradiance survives at $\eta\to\infty$, but finite density, we consider the expansion of Eq.~\eqref{eq:beyond_gamma_B}, to obtain for the instantaneous intensity
\begin{equation}
\label{eq:beyond_bosons_instantaneous_thermodynamic}
\lim_{\eta\rightarrow \infty}\frac{I_\text{B}}{I_\text{dist.}} = 1 + \Ng\frac{3\sqrt{\pi}}{8\eta}\,.
\end{equation}
Thus, bosonic enhancement remains non-vanishing in the thermodynamic limit if the density $\Ng/\eta$ is held constant. This implies that while the global superradiant scaling $\Ne\Ng$, which is characteristic of the Dicke limit, requires infinite density, a weakened form of collective emission survives at finite density $\Nm/\eta$ (number of particles within a single optical wavelength). In this regime, the radiated intensity scales linearly with the particle number but remains enhanced by the local density, reflecting a transition where collective effects are no longer governed by the total ensemble size, but by the number of indistinguishable particles per optical wavelength.

\begin{figure*}[!t]
    \centering
    \includegraphics[]{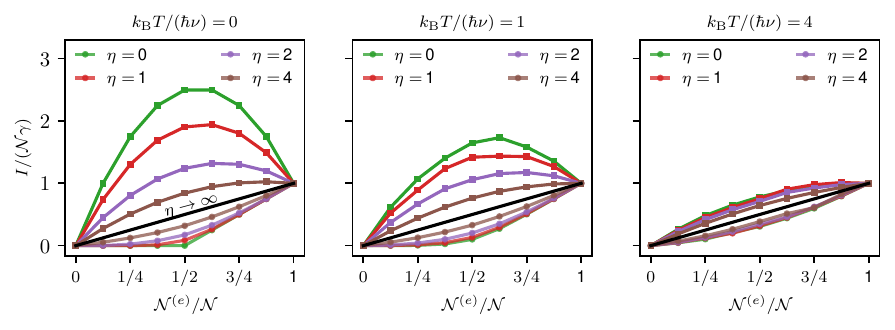}
    \caption{\textbf{Weakening of statistical effects on instantaneous intensity in soft traps and at finite temperature.} Instantaneous intensity from $\Nm=8$ bosons ($\square$) and fermions ($\bigcirc$) as a function of the excitation fraction $\Ne/\Nm$ for different trap widths $\eta$ and different temperatures. At $T=0$ as $\eta$ is increased, the instantaneous intensity approaches the intensity of distinguishable particles $\gamma \Ne$, indicated with the linear black curve. Bosons approach this line from above, while fermions approach it from below. When the temperature is increased, the effects of particle statistics diminish already for smaller trap widths, becoming negligible at high temperatures.}
    \label{fig:fig7}
\end{figure*}

At finite temperature, it is no longer possible to give a simple analytic expression for the instantaneous intensity. However, from the diagonal decay rates derived in App.~\label{App:beyond_instantaneous} we can numerically calculate the instantaneous intensity for moderate trap widths, particle numbers, and temperatures. In Fig.~\ref{fig:fig7} we show how both trap width and temperature affect the instantaneous intensity.

\textbf{Fermions: } For fermions at $T=0$ we can write the instantaneous intensity as
\begin{equation}
I_{\text{F}}(T=0) = \Ne\left(\gamma - \gamma_\textrm{F}(\eta, \Ng)\right)\,,
\end{equation}
where $\gamma_\textrm{F}(\eta,\Ng)$ encodes the reduced intensity due to Pauli-blocking, which depends on the trap width and the number of ground-type particles. It is not possible to give a simple closed-form expression for $\gamma_\textrm{F}(\eta,\Ng)$. From the tight-trap regime, we know that for $\eta\rightarrow0$ and $\Ne \leq \Ng$ results in complete blocking of any decay processes, and hence $\gamma_\text{F}(\eta=0, \Ne \leq \Ng) = 1$. In the limit $\eta\rightarrow\infty$ and $\Ng$ finite, the fraction of blocked decay processes vanishes and thus $\gamma_\text{F}(\eta\rightarrow\infty,\Ng) = 0$. This behavior is illustrated in Fig.~\ref{fig:fig7}, where the instantaneous intensity approaches the intensity of independent particles $\gamma\Ne$, as $\eta\rightarrow \infty$. \\
As for bosons, we are interested in whether Pauli-blocking survives the thermodynamic limit $\Nm\rightarrow\infty$ at a fixed density and finite excitation fraction $\Ne/\Ng = \text{const.}$. Since fermions fill the trap levels according to the Pauli-exclusion principle, constant density requires $\sqrt{\Nm}/z_{\text{zpm}}=\text{const.}$, as the spatial extent of the wavefunctions also scales as $\sqrt{\Nm}$. Calculating the instantaneous intensity from the dissipative matrix elements and occupations is challenging. We give a rough calculation based on the exact matrix elements in the appendix [see App.\,\ref{App:instantaneous_intensity_fermions}] However, for $\Nm\rightarrow \infty$ or $\nu \rightarrow 0$, we can find the qualitative behavior from the following simple calculation. At $T=0$ all states up to the Fermi momentum $\hbar k_\text{F}^{(\alpha)} = \sqrt{2m_0 E_\textrm{F}^{(\alpha)}}$ are occupied (the superscript $\alpha$ labeling the two species). This can be rewritten in terms of the zero-point trap width $z_\text{zpm} = \sqrt{\hbar/(2 m_0 \nu)}$ and the Fermi energy $E_\text{F}^{(\alpha)}=\hbar\nu\Nm^{(\alpha)}$. Thus, the Fermi-vector scales as $k_\textrm{F}^{(\alpha)} =\sqrt{\Nm^{(\alpha)}}/z_{\text{zpm}}$, while the recoil from spontaneous emission $k_0$ is constant. The question is now for which fraction of excited particles the momentum $k_0$ large enough to reach a momentum state above the Fermi surface of the ground particles $k_\text{F}^{(g)}$. This is the usual approach to Fermi-blocking for quasi-free Fermi gases, where the momentum is a good quantum number. In a harmonic trap, this only holds for $\Nm\rightarrow\infty$ or $\nu \rightarrow 0$ ($\eta\to\infty$), because the harmonic oscillator eigenstates are not momentum eigenstates. Nevertheless, for large $n$, the harmonic oscillator eigenfunctions $\phi_n$ are well approximated by standing waves. For complete suppression of spontaneous emission, we get the condition
\begin{equation}
\begin{split}
    k_\textrm{F}^{(e)} + k_0 &\leq k_\textrm{F}^{(g)}\\
    \sqrt{\Ne} + \eta &\leq\sqrt{\Ng}\,.
\end{split}
\end{equation}
In this case, there exists no particle in the excited Fermi sea, for which the recoil momentum $k_0$ is large enough to reach above the Fermi level of the ground state particles, resulting in complete suppression of spontaneous emission.
In contrast, suppression of spontaneous emission due to blocking vanishes when
\begin{equation}
\begin{split}
     k_\textrm{F}^{(g)} &\leq k_0\\
    \sqrt{\Ng}&\leq \eta\,.
\end{split}
\end{equation}
In this case, every particle in the excited Fermi sea can reach above the ground-state Fermi level under photon recoil, such that Pauli-blocking vanishes. 

\subsection{Burst analysis and the superradiant boundary}
\begin{figure*}[!t]
    \centering
    \includegraphics[width = \linewidth]{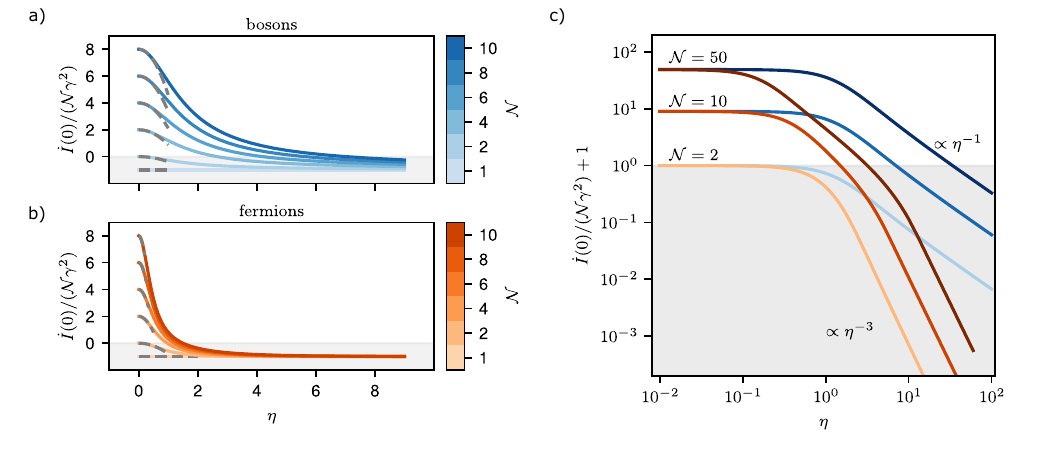}
    \caption{\textbf{Superradiant burst boundary and universal power-law scaling.} a) Initial slope of the intensity for bosons at $T=0$. For $\Nm=1$, the slope is always negative. For $\Nm>2$, the slope is positive for sufficiently small $\eta$, but becomes negative when $\eta$ is increased. Shaded (non-shaded) regions indicate the absence (presence) of a superradiant burst, determined by the sign of the initial slope. b) Same as a), but for fermions. Here, the initial slope decreases more rapidly compared to bosons as the trap width is increased. The dashed grey lines indicate the $\eta \rightarrow 0$ limit from the main text. c) Shifted slope of the intensity on a double-logarithmic plot illustrating the power-law decay of the slope for $\eta \rightarrow \infty$ with bosons in blues and fermions in oranges.}
    \label{fig:fig8}
\end{figure*}
To determine whether a superradiant peak develops without the need for explicit simulation of the dynamics, we determine the derivative of the intensity at $t=0$
\begin{equation}
    \dot I(0) = \sum_{\substack{n,n' \\ m,m'}}\Gamma_{n,n'}^{m,m'}\left(\expval{\dot{\hat{\Sigma}}^\dag_{n,n'}\Sig{m,m'}} + \expval{\Sigd{n,n'}\dot{\hat{\Sigma}}^{\phantom\dag}_{m,m'}}\right)\,,
    \label{eq:beyond_slope_general}
\end{equation}
whose sign predicts the buildup of superradiant burst correlations, indicating no superradiance for $\dot I(0)<0$ and indicating superradiance for $\dot I(0)>0$. The initial slope starting from full excitation can also be related to the second-order correlation function, where photon bunching predicts the buildup of a superradiant burst~\cite{masson2022universality}. Eq.~\eqref{eq:beyond_slope_general} can be evaluated quite efficiently [for details see App.~\ref{App:beyond_intensity}].

 For arbitrary initial inversions, we explicitly calculate the derivative of the intensity at $t=0$ [see App.~\ref{App:beyond_intensity}]. Superradiant decay is characterized by an initial increase in the radiated intensity, since the emission of the first photon results in an enhancement of the emission probability for the second photon. In the Dicke limit with full initial inversion, the initial slope is positive and scales quadratically with the particle number $\dot{I}(0) = +\gamma^2\Nm(\Nm-1)$. In contrast, for independent particles, the intensity decreases monotonically with an initial slope that is negative and scales linearly with the particle number $\dot I(0)=-\gamma^2\Nm$. We derive in Appendix~\ref{App:beyond_intensity} general expressions for the initial slope of the intensity. Because bosons at $T=0$ only occupy the ground state of the trap, it is possible to derive compact analytical expressions for the initial slope of the intensity. 

\textbf{Bosons: } For bosons at $T=0$, the required matrix elements simplify significantly, and we can give relatively compact expressions for the initial slope of the intensity  [see App.~\ref{App:beyond_intensity}]. If we additionally assume $\Ne = \Nm, \Ng = 0$, we obtain the limiting cases
\begin{equation}
    \label{eq:beyond_bosons_slope_thermodynamic}
        \frac{\dot I_\text{B}(0)}{\Nm\gamma^2} =
    \begin{cases}
        (\Nm-1)\left(1 - \dfrac{17}{45}\eta^2\right) -1 &\eta \rightarrow 0 \\
         \dfrac{3\sqrt{\pi}}{8\eta}(\Nm -1 ) - 1 &\eta \rightarrow \infty\,.
    \end{cases}
\end{equation}
Both expressions are exact to leading order and are derived analytically. The derivation and more general expressions are given in the appendix [see App.~\ref{App:beyond_intensity}]. In Fig.~\ref{fig:fig8}a) we compare the analytical expressions to numerical simulations, showing good agreement in the relevant regimes. Therefore, as was the case with the instantaneous intensity, particle statistics effects remain in the thermodynamic limit at constant density, albeit in reduced form. Further, the scaling of the slope agrees with the scaling of the peak intensity in the Lamb-Dicke regime.\\
In the wide trap limit, both Eq.~\eqref{eq:beyond_bosons_instantaneous_thermodynamic} and Eq.~\eqref{eq:beyond_bosons_slope_thermodynamic} depend only on the density $\Nm/\eta = \Nm\lambda_0/(2\pi z_\textrm{zpm})$. An ensemble of $\Nm$ bosonic particles at $T=0$ in a harmonic trap has an approximate spatial extent given by the zero-point trap with $z_\text{zpm}$. If the trap is sufficiently narrow or the number of particles sufficiently high, such that there is more than one indistinguishable particle per optical wavelength, superradiance survives the thermodynamic limit in a weakened form, depending only on the number of particles per optical wavelength, not on the total number of particles.

\textbf{Fermions: } For fermions, it is not possible to give a simple analytical expression for the initial slope of the intensity, even for $T=0, \Ng = 0$. However, numerically, we can evaluate the initial slope for fermions for a broad range of particle numbers, trap widths, and temperatures. In Fig.~\ref{fig:fig8}\,b) we show the initial slope of the intensity for different numbers of particles as a function of the trap width at $T=0$. Compared to bosons, the initial slope of the intensity for fermions reduces more rapidly as the trap width is increased. This aligns with the previous results in the Lamb-Dicke regime [see Fig.~\ref{fig:fig5}c)-f)]. Because even at $T=0$ fermions occupy the first $\Nm$ trap levels, rather than just the ground state in the case of bosons, the permutational symmetric component decreases more rapidly as the trap width is increased, resulting in a decrease $\propto -\Nm\eta^2$ for fermions, rather than the $\propto-\eta^2$ decrease for bosons, which is independent of particle number.
From the numerical calculation, we find the initial slope of the intensity for fermions in the two limiting cases
\begin{equation}
    \frac{\dot I_\text{F}(0)}{\Nm\gamma^2} \approx
    \begin{cases}
        (\Nm-1)\left(1 - \alpha\Nm\eta^2\right)-1 &\eta \rightarrow 0 \\
         \dfrac{\beta}{\eta^3}(\Nm -1 ) - 1 &\eta \rightarrow \infty\,,
    \end{cases}
\end{equation}
where $\alpha\approx 0.4$ and  $\beta = \mathcal{O}(1)$. The exact value of $\beta$ is hard to pin down because fermions occupy higher trap levels, resulting in highly oscillatory integrals. As for bosons, the narrow trap limit agrees well with the burst scaling from the Lamb-Dicke regime, resulting in a decrease of the peak intensity with $-\Nm\eta^2$, for fermions, due to the larger spatial extent, and hence higher distinguishability of higher trap states. For wide traps, this also results in the steeper power-law decay $\propto \eta^{-3}$. Thus, for fermions in the thermodynamic limit, no superradiant burst occurs at finite density in this setting.

\section{Conclusions and outlook}
\subsection{Conclusions}
We have investigated the purely dissipative dynamics of degenerate quantum gases in a 1-dimensional harmonic trap and have shown the interplay between particle statistics and collective radiance. While the purely dissipative model is somewhat idealized, it has allowed us to reveal a rich interplay between collective radiance and particle statistics, which derive from the permutational symmetry of the Lindbladian and the symmetry of the exchange of indistinguishable particles. We have also shown how the usual spin-based description of distinguishable particles in standard quantum optics emerges from the more general field-theoretic description.\\ 
In the \emph{tight-trap regime}, the Lindbladian is completely permutationally symmetric, corresponding to the usual Dicke limit. In this regime, particle statistics enter only through the decomposition of the initial state into collective spin components. While bosonic symmetrization results in larger collective spin values compared to distinguishable particles, fermionic antisymmetrization results in smaller collective spin values. Furthermore, we have shown how, as the temperature is increased, the effects of particle statistics diminish.\\
In the \emph{Lamb-Dicke regime}, the permutational symmetry of the Lindbladian is reduced, and additional decay channels emerge, where decay includes changes in the trap state. The reduction of permutational symmetry is accompanied by a reduction in the peak intensity of the superradiant burst at $T=0$, which is more pronounced for fermions compared to bosons. Further, the reduction in permutational symmetry is accompanied by long-range transport of particles in the trap and long sub-radiant tails mediated by redistribution of excitations induced by the non-Hermitian Hamiltonian.\\
\emph{Beyond the Lamb-Dicke regime}, we derived scaling laws for the superradiant dynamics as a function of the trap width. We found that for large trap widths, the initial slope of the intensity follows a power-law decay with exponents of $-1$ and $-3$ for bosons and fermions, respectively. Crucially, these exponents are independent of the particle number $\Nm$, identifying them as universal signatures of the interplay between quantum statistics and the spatial structure of the harmonic trap eigenfunctions.

\subsection{Outlook}\label{sec6}
There are numerous intriguing extensions and variations of the scenario considered in the paper. Further, some of the methods applied here can be easily applied to variations and extensions of the model considered.

In particular, we plan to further investigate the interplay of collective radiance, particle statistics, and recoil-induced motion. We believe that the following scenarios are promising. First, one- and two-dimensional scenarios in infinitely deep box potentials, where the dissipative matrix elements have a simple structure due to momentum conservation. Second, cavity and waveguide scenarios, where only a few collective decay channels are present.

While applied here to photon recoil, our perturbative basis generation is applicable to any system where a global symmetry-protected manifold is broken by local terms. This includes local dephasing in collective ensembles or short-range density-density interactions in waveguide QED. This forms a bridge to facilitate the analysis of the dynamics in nearly permutationally invariant systems in a numerically efficient manner.

To make the treatment more realistic, there are various aspects that were not treated here. One is the robustness of the observed effects when including s-wave scattering and dephasing. Furthermore, the role of unitary interactions for the dissipative dynamics discussed here is still unclear, and the degeneracy of the harmonic oscillator in higher dimensions is expected to lead to qualitatively different physics and scaling laws.

\section*{Acknowledgments}
We acknowledge fruitful discussions with A. Jaber. Numerical simulations were performed in Julia~\cite{Julia-2017}, using the packages DifferentialEquations.jl~\cite{Rackauckas2017}, QuantumOptics.jl~\cite{kramer2018quantumoptics}, Matplotlib~\cite{Hunter:2007}, ExponentialUitilities.jl, FastGaussQuadratures.jl, PartialWavefunctions.jl, ClassicalOrthogonalPolynomials.jl, Einsum.jl. We acknowledge financial support from the Max Planck Society and from the Deutsche Forschungsgemeinschaft (DFG, German Research Foundation) -- Project-ID 429529648 -- TRR 306 QuCoLiMa ("Quantum Cooperativity of Light and Matter’’).

\FloatBarrier
\bibliography{bibliography}
\onecolumngrid

\appendix
\newpage
\section{Elimination of electromagnetic vacuum in the trap basis}
\label{App:Elimination}
In this section of the appendix we perform the elimination of the electromagnetic vacuum starting from a field-theoretic description of the dipolar light-matter Hamiltonian under the usual Born and Markov approximation. \\
In order to derive the field-theoretic Lindblad master equation, we start from the following light-matter Hamiltonian describing dipolar emitters coupled to the electromagnetic vacuum in a field-theoretic formulation~\cite{lewenstein1994quantum}
\begin{equation}
\label{eq:field_theory}
    \begin{split}
     H_a&=\int\diff{\bm R}\left[\omega_0+\frac{\bm P^2}{2M}+V(\bm R)\right]\Psi_e^\dagger(\bm R)\Psi_e(\bm R)+\left[\frac{\bm P^2}{2M}+V(\bm R)\right]\Psi_g^\dagger(\bm R)\Psi_g(\bm R)\\
     H_{lm}&=\int\diff{\bm R}\Big[\left(\Psi_g^\dagger(\bm R)\Psi_e(\bm R)\bm d  + \Psi_g(\bm R)\Psi_e^\dagger(\bm R)\bm d^* \right)\hat{\bm E}(\bm R)\Big]\,.
    \end{split}
\end{equation}
where we used the definition of the electric-field operator
\begin{equation}
\hat{\bm E}(\bm R)=\iu\sum_{\bm k,\bm\epsilon}\sqrt{\frac{\omega_k}{2V\epsilon_0}}\left[\bm\epsilon_{\bm k}^* \me^{\iu\bm k\bm R}a_{\bm k.\bm\epsilon_{\bm k}}\,-\bm\epsilon_{\bm k} \me^{-\iu\bm k\bm R}a^\dagger_{\bm k.\bm\epsilon_{\bm k}}\right].
\end{equation}
Transforming into the trap basis, where we denote the harmonic oscillator eigenfunctions as $\phi_{\bm m}(\bm R)$, diagonalizes the atomic Hamiltonian $H_a$. We thus obtain in the trap basis
\begin{equation}
\label{eq:field_theory_trap_basis}
    \begin{aligned}
     H_a&=\sum_{\bm n}\left[(\tf_{\bm n}+\omega_0)\hat e_{\bm n}^\dagger\hat e_{\bm n}+\tf_{\bm n}\hat g_{\bm n}^\dagger\hat g_{\bm n}\right],\\
     H_{lm}&=\sum_{\bm n,\bm n'}
     (\bm d \gd{\bm n} \e{\bm n'}+\bm d^* \ed{\bm n} \g{\bm n'})\left[\int\diff{\bm R}\phi_{\bm n}(\bm R)\phi^*_{\bm n'}(\bm R)\hat{\bm E}(\bm R)\right]\,.
    \end{aligned}
\end{equation}

We can thus identify a system-bath coupling operator
\[
\mathcal F_{\bm n,\bm n'}=\int\diff{\bm R}\phi_{\bm n}(\bm R)\phi^*_{\bm n'}(\bm R)\bm d^*\cdot \hat{\bm E}(\bm R),
\]
such that the light-matter hamiltonian taks the form
\[
H_{lm}=\sum_{\bm n,\bm n'}(\mathcal F_{\bm n,\bm n'}^\dagger\gd{\bm n} \e{\bm n'}+\mathcal F_{\bm n,\bm n'}\ed{\bm n'} \g{\bm n}).
\]

The relevant bath correlation function is then
\begin{equation}
   \begin{aligned}
    C_{\bm n,\bm n'}^{\bm m,\bm m'}(\tau)&=\expval{\mathcal F_{\bm n,\bm n'}(\tau)\mathcal F^\dagger_{\bm m,\bm m'}(0)}\\
    &=\int\diff{\bm R}\diff{\bm R'}\int\phi_{\bm n}(\bm R)\phi^*_{\bm n'}(\bm R)
    \phi_{\bm m'}(\bm R')\phi^*_{\bm m}(\bm R')\expval{\bm d^*\cdot \hat{\bm E}(\bm R,t)\bm d\cdot \hat{\bm E}(\bm R',0)}.
\end{aligned} 
\end{equation}

The electric field correlation function can now be obtained in a standard treatment, using $\expval{a_{\bm k',\bm\epsilon}(t)a^\dagger_{\bm k',\bm\epsilon'}(0)}=\delta_{\bm k,\bm k'}\delta_{\bm\epsilon,\bm\epsilon'}\me^{-\iu\omega_k t}$ and $\expval{a^\dagger_{\bm k',\bm\epsilon}a_{\bm k',\bm\epsilon'}}\approx 0$
\[
\expval{\hat{\bm E}(\bm R,t)\otimes \hat{\bm E}(\bm R',0)}=\sum_{\bm k,\bm\epsilon}\frac{\omega_k}{2V\epsilon_0}\epsilon^*_{\bm k}\otimes \epsilon_{\bm k}\me^{\iu(\bm R-\bm R')\bm k}\me^{-\iu\omega_k t}.
\] 

This correlation function now needs to be evaluated at the frequency of the operator it is paired with, i.e. $\ed{\bm n'}\g{\bm n}$ and $\gd{\bm n}\e{\bm n'}$. These operators rotate with a frequency $\omega_0+\nu_{\bm n'}-\nu_{\bm n}\approx\omega_0$ and $-(\omega_0+\nu_{\bm n'}-\nu_{\bm n})\approx -\omega_0$ respectively. We thus obtain the correlation functions
\[
\expval{\hat{\bm E}(\bm R)\otimes \hat{\bm E}(\bm R')}(\pm\omega_0)=\lim_{t\to\infty}\int_{0}^t\diff{\tau}\sum_{\bm k,\bm\epsilon}\frac{\omega_k}{2V\epsilon_0}\epsilon^*_{\bm k}\otimes \epsilon_{\bm k}\me^{\iu(\bm R-\bm R')\bm k}\me^{\iu(\pm\omega_0-\omega_k)\tau},
\]
where the limit $t\to\infty$ constitutes the Markovian approximation and $\otimes$ denotes the tensor product on the three dimensional cartesian axis vector space. The Fourier transform of a plane-wave in the distributional sense is given as
\[
\lim_{t\to\infty}\int_{0}^t\diff{\tau}\me^{\iu(\pm\omega_0-\omega_k)\tau}=\iu\mathcal P\frac{1}{\pm\omega_0-\omega_k}+\pi\delta(\pm\omega_0-\omega_k).
\]

From the orthonomality of the wave vector and the polarizations one obtains
\[
\sum_{\bm\epsilon}\epsilon^*_{\bm k}\otimes \epsilon_{\bm k}=\mathds 1-\hat k\otimes \hat k,
\]
where $\hat k$ is the unit vector in direction $\bm k$. At the same time we take the continuum limit for the wavevectors
\[
\sum_{\mathbf k,\bm\epsilon}\longrightarrow\frac{V}{(2\pi)^3}\int \diff{\bm k}\sum_{\bm\epsilon}.
\]

Putting this together, the bath correlation function for the electric field operators becomes
\[\begin{aligned}
    \mathcal C_E(\omega_0)&=\expval{\hat{\bm E}(\bm R)\otimes \hat{\bm E}(\bm R')}(\omega_0)=\int \diff{\bm k}\frac{\omega_k}{2\epsilon_0(2\pi)^3}\left(\mathds 1-\hat k\otimes \hat k\right)\left[\iu\mathcal P\frac{1}{\omega_0-\omega_k}+\pi\delta(\pm\omega_0-\omega_k)\right]\me^{\iu(\bm R-\bm R')\bm k}\\
    \mathcal C_E(-\omega_0)&=\expval{\hat{\bm E}(\bm R)\otimes \hat{\bm E}(\bm R')}(-\omega_0)=\int \diff{\bm k}\frac{\omega_k}{2\epsilon_0(2\pi)^3}\left(\mathds 1-\hat k\otimes \hat k\right)\left[\iu\mathcal P\frac{1}{-\omega_0-\omega_k}\right]\me^{\iu(\bm R-\bm R')\bm k},
\end{aligned}
\]
since $\delta(-\omega_0-\omega_k)=0$. 

This leads to the canonical definition of the free-space Helmholtz Green's function in three dimension in Fourier space given by
\begin{equation}
\widetilde{\mathbf G}(\bm k)=\frac{\mathds 1-\hat k\otimes \hat k}{k_0^2-k^2+\iu 0^+}=\frac{1}{2k_0} \left[\frac{2 k_0}{k_0^2 - k^2} -\iu \pi \delta(k_0 - k)\right]\left(\bm 1 -\bm \hat{k}\otimes \bm \hat{k}\right).
\end{equation}

The Fourier transform of the Green's function is then given by
\begin{equation}
\mathbf G(\bm R)=\frac{1}{(2\pi)^3}\int\diff{\bm k}\me^{-\iu\bm k\cdot\bm R}\widetilde{\mathbf{G}}(\mathbf{k})=\left(\bm 1 + \frac{\bm\nabla \otimes \bm\nabla}{k_0^2} \right)\frac{\me^{\iu k_0  R}}{4\pi R} - \frac{\bm 1}{3k_0^2}\delta(\bm R).
\end{equation}

The sum of the positive and negative frequency contributions to the electric field correlator can then be written in terms of these Green's functions as
\[\begin{aligned}
    \mathcal C_E(\omega_0)+\mathcal C_E(-\omega_0)&=\iu\int \diff{\bm k}\frac{k^2}{2\epsilon_0(2\pi)^3}\widetilde{\mathbf G}(\bm k)\me^{\iu(\bm R-\bm R')\bm k}
    =-\iu\frac{1}{2\epsilon_0}\nabla^2 G(\bm R-\bm R')=\iu\frac{k_0^2}{2\epsilon_0}G(\bm R-\bm R').
\end{aligned}
\]
The last line only strictly holds at $\bm R\neq \bm R'$. Using $\mu_0\epsilon_0=c^2$ we thus recover the definition in the main text
\[\begin{aligned}
   J_{\bm n,\bm n'}^{\bm m,\bm m'} &= -\mu_0 \omega_0^2\Re\left[ \bm d^\dag \mathbf{G}_{\bm n,\bm n'}^{\bm m,\bm m'}\bm d \right]\\
	\Gamma_{\bm n,\bm n'}^{\bm m,\bm m'} &= 2\mu_0 \omega_0^2\Im\left[\bm d^\dag \mathbf{G}_{\bm n,\bm n'}^{\bm m,\bm m'} \bm d \right], 
\end{aligned}
\]
where we defined
\begin{equation}
\begin{aligned}
	\mathbf{G}_{\bm n,\bm n'}^{\bm m,\bm m'} &=\int\diff{\bm R}\diff{\bm R'}\Phi_{\bm n,\bm n'}(\bm R)\Phi_{\bm m,\bm m'}(\bm R')\mathbf{G}(\bm R-\bm R')\\
    &=\frac{1}{(2\pi)^3}\int\diff{\bm k}\eta_{\bm n, \bm n'}^*(\bm k)\eta_{\bm m,\bm m'}(\bm k)\widetilde{\mathbf{G}}(\bm k),
\end{aligned}
\end{equation}
with $\Phi_{\bm n,\bm n'}(\bm R)=\phi^*_{\bm n}(\bm R)\phi_{\bm n'}(\bm R)$.

\section{Properties of associated Laguerre Polynomials}

The associated Laguerre Polynomials are defined as
\begin{equation}
L_n^{(m)}(x)=\sum_{i=0}^n (-1)^i {n+m \choose n-i} \frac{x^i}{i!}=(n+m)!\sum_{i=0}^n (-1)^i \frac{x^i}{(m+i)!(n-i)!i!},
\end{equation}
so that
\[
\begin{aligned}
    L_n^{(n)}(x)&=\frac{x^n}{n!}\\
    L_n^{(0)}(x)&=L_n(x).
\end{aligned}
\]

\section{Spin decomposition in the tight trap regime}\label{App:spin_decomposition}
In this section we give some details on the spin decomposition in the tight trap regime. We begin by defining the following collective operators
\begin{equation}
   \begin{aligned}
    \Sig{0}{(z)}&=\sum_{n=0}^\infty\left(\ed{n} \e{n}-\gd{n} \g{n}\right)/2,\\
    \Sig{0}{(x)}&=(\Sigd{0}+\Sig{0})/2,\\
    \Sig{0}{(y)}&=(\Sigd{0}-\Sig{0})/(2\iu),
\end{aligned} 
\end{equation}
which obey the $\text{SU}(2)$-spin algebra $\commutator{\Sig{0}{(j)}}{\Sig{0}{(k)}} = \iu \varepsilon_{jkl}\Sig{0}{(l)}$, with $\varepsilon_{jkl}$ the Levi-Civita symbol. The local operators on each of the trap levels
\begin{align}
    \hat{s}^{(z)}_n&=(\hat{e}^\dag_n \hat{e}^{\phantom{\dagger}}_n - \hat{g}^\dag_n \hat{g}^{\phantom{\dagger}}_n)/2,\\
    \hat{s}^{(x)}_n&=(\hat{e}^\dag_n \hat{g}^{\phantom{\dagger}}_n + \hat{g}^\dag_n \hat{e}^{\phantom{\dagger}}_n)/2,\\
    \hat{s}^{(y)}_n&=(\hat{e}^\dag_n \hat{g}^{\phantom{\dagger}}_n - \hat{g}^\dag_n \hat{e}^{\phantom{\dagger}}_n)/(2\iu),
\end{align}
also obey the $\text{SU}(2)$-spin algebra.

This motivates a description in terms of a spin degree of freedom encoded on each trap level. Effectively, this replaces the two quantum numbers $\Ne_n, \Ng_n$ with the new quantum numbers $s_n,m_n$ (spin and magnetization). For bosons, this mapping corresponds to the inverse of Schwinger-Bosonization~\cite{schwinger1952angular,sakurai2020modern}, and for fermions it is inverse Abrikosov fermionization \cite{Abrikosov1965electron}
\begin{equation}
\begin{split}
        s_{n} &= 
        \begin{cases} 
        \left(\Nm^{(e)}_n + \Nm^{(g)}_n\right) /2 \quad &\text{bosons} \\[2ex]
        \left| \Nm^{(e)}_n - \Nm^{(g)}_n \right| /2 \quad &\text{fermions} \\
        \end{cases} \\[1ex]
        m_{n} &= \left(\Nm^{(e)}_n - \Nm^{(g)}_n\right)/2.
\end{split}
\end{equation}

The conservation of particles on trap-level $n$ is then translated to the conservation of the spin $s_n$. We can thus understand the global spin as the spin addition of the local spins $\{\hat{\bm s}_0,\hat{\bm s}_1,\ldots\}$ as $\hat{\bm S}_0=\sum_{n=0}^\infty \hat{\bm s}_n$. Here we defined the spin vectors $\hat{\bm S}_0=(\Sig{0}{(x)},\Sig{0}{(y)},\Sig{0}{(z)})^T$ and $\hat{\bm s}_n=(\hat{s}^{(x)}_n,\hat{s}^{(y)}_n,\hat{s}^{(z)}_n)^T$. The total Hilbert $\Hm_\text{tot}$ constructed from the tensor product of the local spaces $\Hm_{s_n}$ with $s_n$ fixed to the level $n$ can thus be decomposed into a direct sum of collective spins
\begin{equation}
    \mathcal{H}_{\text{tot}} = \bigotimes_{n=0}^\infty \mathcal{H}_{s_n} \cong \bigoplus_{S=0}^{\Nm/2} \bigoplus_{\alpha=1}^{d_S} \mathcal{H}^{(\alpha)}_S,
\end{equation}
where $d_S$ is the multiplicity of each collective spin sector and $\alpha$ is the multiplicity quantum number.  The degeneracies of the collective spin subspaces, when constructed from $\Nm$ spin-1/2 Hilbert spaces, were given by Dicke~\cite{Dicke54}. The generalization to spins of arbitrary sizes is given in Ref.~\cite{zborovsky1981addition}. In the usual setting of Dicke superradiance, only the nondegenerate space $S=N/2$ is occupied. Every state $\ket{\psi}\in\mathcal{H}_{\text{tot}}$ admits a representation
\begin{equation}
\ket{\psi}=\sum_{S=0}^{\Nm/2}\sum_{M=-S}^S\sum_{\alpha=1}^{d_S}c_{S,M,\alpha}\ket{S,M,\alpha}.
\end{equation}

Crucially, the action of collective operators is independent of the quantum number $\alpha$, for instance, the action of the collective collapse operator is given by

\begin{equation}
\Sig{0}\ket{S,M,\alpha}=\sqrt{S(S+1)-M(M-1)}\ket{S,M-1,\alpha}.
\end{equation}

Therefore, collective observables for some collective spin $S$ can be computed independently of $\alpha$. If we consider an initial product state with fixed $M$, the resulting density matrix becomes

\begin{equation}
\rho=\sum_{S,S'=0}^{\Nm/2}\sum_{\alpha,\alpha'=1}^{d_S}c_{S,M,\alpha}c_{S',M,\alpha'}\outerproduct{S,M,\alpha}{S',M,\alpha'}.
\end{equation}

In practice, however, the dynamics of the coherences and the populations decouple in the tight trap regime. Since only the populations are required to calculate the intensity, we can thus consider the projection of the density matrix onto its populations
\begin{equation}
\rho_{\text{diag}}=\sum_{S=0}^{\Nm/2}\sum_{\alpha=1}^{d_S}p_{S,M,\alpha}\outerproduct{S,M,\alpha}{S,M,\alpha}.
\end{equation}
with $p_{S,M,\alpha}=\abs{c_{S,M,\alpha}}^2$. Then we proceed as discussed in the main text. We trace over the degeneracy quantum number, as collective observables do not depend on it, and derive the reduced populations $p_{S,M}$. This reduces the dimension of the effective Hilbert space to $\propto \Nm^2$.

\section{Matrix elements of the dipolar Green's function in the trap basis}\label{App:matrix_eta}

A closed form for the Franck-Condon factors may be found by seeing that the matrix elements
\[
\eta_{n,m}(\bm k')=\matrixelement{\bm n}{\me^{-\iu k_z \hat z}}{\bm m}=\matrixelement{\bm n}{\mathcal D[-\iu k_z z_{\rm{zpm}}/\sqrt{2}]}{\bm m},
\]
where $\mathcal D[\alpha]=\me^{\alpha^* \hat a-\alpha \hat a^\dagger}$ is the displacement operator for the harmonic oscillator for the $z$-component. The matrix elements of displacement operators in the Fock basis are known~\cite{cahill1969ordered}

\begin{equation}
\eta_{n,m}(k) = \sqrt{\frac{m!}{n!}}\left( \frac{-\iu z_{\rm{zpm}} k}{\sqrt{2}} \right)^{n - m}\me^{- \frac{z_{\rm{zpm}}^2 k^2}{4}}L_{m}^{(n - m)}\left( \frac{z_{\rm{zpm}}^2 k^2}{2} \right).
\end{equation}

The vectorial expression factorizes into its cartesian components

\[
\eta_{\bm n,\bm  m}(\bm{k})=\eta_{n_x,m_x}(k_x)\eta_{n_y,m_y}(k_y)\eta_{n_z,m_z}(k_z).
\].

This representation can be used to find the matrix elements of the Green's function in the trap basis. We shall assume, as in the main text, that $n_x=m_x=n_y=m_y=n'_x=m'_x=n'_y=m'_y=0$. We then set $n_z=n,n'_z=n',m_z=m,m'_z=m$ to The relevant matrix element then
\begin{equation}
   \begin{aligned}
    \mathbf{G}^{m,m'}_{n,n'}&=\frac{(-1)^{n-n'}}{(2\pi)^3}\sqrt{\frac{n'!m'!}{n!m!}}\int\diff{\bm k}\exp\left[- \frac{x_{\rm{zpm}}^2 k_x^2}{2}- \frac{y_{\rm{zpm}}^2 k_y^2}{2}- \frac{z_{\rm{zpm}}^2 k^2}{2}\right]
    \left(\frac{\iu z_{\rm{zpm}} k_z}{\sqrt{2}} \right)^{n+m-n'-m'}\\
    &\times L_{n'}^{(n - n')}\left( \frac{z_{\rm{zpm}}^2 k_z^2}{2} \right)L_{m'}^{(m - m')}\left( \frac{z_{\rm{zpm}}^2 k_z^2}{2} \right)
    \left( \mathbb{I} - \hat{\mathbf{k}} \otimes \hat{\mathbf{k}} \right)\left[ \mathcal{P} \frac{2k_0}{k_0^2 - k^2} - \iu \pi \delta(k - k_0) \right].
\end{aligned} 
\end{equation}

We thus introduce $u=k/k_0$ and $\eta_x=k_0x_{\rm{zmp}}/\sqrt{2}$,$\eta_y=k_0y_{\rm{zmp}}/\sqrt{2}$ and $\eta_z=k_0z_{\rm{zmp}}/\sqrt{2}$ to substitute in the integration and transform to spherical coordinates to obtain

\[
\begin{aligned}
    \mathbf{G}^{m,m'}_{n,n'}&=\frac{(-1)^{n-n'}}{(2\pi)^3}k_0^2\sqrt{\frac{n'!m'!}{n!m!}}\iu^{n+m-n'-m'}\int\diff{u}\diff{\cos(\theta)}\diff{\phi}u^2
    \me^{- u^2\left(\eta_x^2\cos^2(\theta)\sin^2(\theta)- \eta_y^2\sin^2(\theta)\sin^2(\theta)-\eta_z^2\cos^2(\theta)\right)}\\
    &\times \left(\eta_zu\cos(\theta)\right)^{n+m-n'-m'}L_{n'}^{(n - n')}\left(\eta_z^2u^2\cos^2(\theta)\right)L_{m'}^{(m - m')}\left(\eta_z^2u^2\cos^2(\theta)\right)
    \bm T(\theta,\phi)\left[ \mathcal{P} \frac{2}{1 - u^2} - \iu \pi \delta(u - 1) \right].
\end{aligned}
\]

where we defined the projector $\bm T(\theta,\phi)$ onto the plane orthogonal to $\mathbf{\hat k}$

\[
\bm T(\theta,\phi)=\mathds 1-\left(
\begin{array}{ccc}
 \sin ^2(\theta ) \cos ^2(\phi ) & \sin ^2(\theta ) \sin (\phi ) \cos (\phi ) & \sin (\theta ) \cos (\theta ) \cos (\phi ) \\
 \sin ^2(\theta ) \sin (\phi ) \cos (\phi ) & \sin ^2(\theta ) \sin ^2(\phi ) & \sin (\theta ) \cos (\theta ) \sin (\phi ) \\
 \sin (\theta ) \cos (\theta ) \cos (\phi ) & \sin (\theta ) \cos (\theta ) \sin (\phi ) & \cos ^2(\theta ) \\
\end{array}
\right).
\]

The key is now the expansion of the associated Laguerre polynomials in terms of monomials $L_n^{(m)}(x)=\sum_{i=0}^n (-1)^i {n+m \choose n-i} \frac{x^i}{i!}$

\[
\begin{aligned}
    \mathbf{G}^{m,m'}_{n,n'}&=\frac{(-1)^{n-n'}}{(2\pi)^3}k_0^2\sqrt{\frac{n'!m'!}{n!m!}}\iu^{n+m-n'-m'}\sum_{i=0}^{n'}\sum_{j=0}^{m'}
    {n \choose n'-i}{m \choose m'-j}\frac{\eta_z^{n+m-n'-m'+2(i+j)}}{i!j!}\\
    &\times\int\diff{u}\diff{\cos(\theta)}\diff{\phi}u^2
    \me^{- u^2\left(\eta_x^2\cos^2(\theta)\sin^2(\theta)- \eta_y^2\sin^2(\theta)\sin^2(\theta)-\eta_z^2\cos^2(\theta)\right)}
    \left(u\cos(\theta)\right)^{n+m-n'-m'+2(i+j)}\\
    &\times\bm T(\theta,\phi)\left[ \mathcal{P} \frac{2}{1 - u^2} - \iu \pi \delta(u - 1) \right].
\end{aligned}
\]

In order to find $\bm \Gamma^{m,m'}_{n,n'}$ we must thus calculate

\[
\bm I_l(\eta_x,\eta_y,\eta_z)=\int\diff{u}\diff{\cos(\theta)}\diff{\phi}u^2
    \me^{- u^2\left(\eta_x^2\cos^2(\theta)\sin^2(\theta)- \eta_y^2\sin^2(\theta)\sin^2(\theta)-\eta_z^2\cos^2(\theta)\right)}
    \left(u\cos(\theta)\right)^{l}
    \bm T(\theta,\phi)\left[\pi \delta(u - 1) \right].
\]

so that, assuming $n_z\geq n_z'$ and $m_z\geq m_z'$ using $\gamma=\frac{k_0^3d^2}{3\pi\epsilon_0}$

\[
\begin{aligned}
    \Gamma^{m,m'}_{n,n'}&=\gamma \frac{3}{8\pi^2}\sqrt{\frac{n'!m'!}{n!m!}}(-1)^{n-n'}\iu^{n+m-n'-m'}\\
    &\times \sum_{i=0}^{n'}\sum_{j=0}^{m'}
    {n \choose n'-i}{m \choose m'-j}\frac{\eta^{n+m-n'-m'+2(i+j)}}{i!j!}(-1)^{i+j}\hat{\bm d}\cdot \bm I_{n+m-n'-m'+2(i+j)}(\eta_x,\eta_y,\eta_z)\cdot \hat{\bm{d}}.
\end{aligned}
\]

We consider for simplicity $\bm I_{l}(0,0,\eta_z)$ which is a diagonal matrix. The general matrix elements can then be found as
\[
\begin{aligned}
    \bm I^{(xx)}_{l}(0,0,\eta_z)&=
    4\pi^2\cos\bigl(\tfrac{\pi}{2}l\bigr)\sum_{j=0}^{\infty}\eta_z^{2j}\frac{(-1)^j}{j!}\frac{l+2j+2}{(l+2j+1)(l+2j+3)} \\
    &= 2\pi^2 \cos\bigl(\tfrac{\pi}{2}l\bigr)\sum_{j=0}^{\infty} \eta^{2j}\frac{(-1)^j}{j!}\left[\frac{1}{2j + l + 1} + \frac{1}{2j + l +3}\right]
    \\
    &= \pi^2\cos\bigl(\tfrac{\pi}{2}l\bigr)\left[\frac{1}{\eta^{{l+1}}}\Gammalower(\tfrac{l+1}{2},\eta^2) + \frac{1}{\eta^{{l+3}}}\Gammalower(\tfrac{l+3}{2},\eta^2)\right]\\
    \bm I^{(zz)}_{l}(0,0,\eta_z)
    &=8\pi^2 \cos\left(\tfrac{\pi}{2}l\right)\sum_{j=0}^\infty \eta^{2j}\frac{(-1)^j}{j!}\frac{1}{(l+2j+1)(l+2j+3)} \\
    &= 4\pi^2\cos\left(\tfrac{\pi}{2}l\right)\sum_{j=0}^{\infty}\eta^{2j}\frac{(-1)^j}{j!}\left[\frac{1}{2j + l + 1} - \frac{1}{2j + l +3}\right]\\
    &= 2\pi^2\cos\bigl(\tfrac{\pi}{2}l\bigr)\left[\frac{1}{\eta^{l+1}}\Gammalower(\tfrac{l+1}{2},\eta^2) - \frac{1}{\eta^{{l+3}}}\Gammalower(\tfrac{l+3}{2},\eta^2)\right]
\end{aligned}
\]
where the superscript denotes the dipolar orientations and with $\bm I^{(yy)}_{l}(0,0,\eta_z)=\bm I^{(xx)}_{l}(0,0,\eta_z)$ by symmetry, since we have chosen the $z-$axis as the axis of motion. From the partial fraction decomposition, we obtain the final result in terms of lower incomplete gamma functions $\Gammalower(.)$. Thus, the general rates can be rewritten as (for $n'\leq n$ and $m'\leq m$) and the dipoles aligned in $z-$ direction $\hat d=\hat e_z$ with $l:= n + m -n' -m' + 2(i+j)$
\begin{equation}
    \label{eq:Appendix_Gammatensor_general}
    \Gamma^{m,m'}_{n,n'}=\gamma(-1)^{n-n'} \frac{3}{4}\sqrt{\frac{n'!m'!}{n!m!}}\iu^{n+m-n'-m'}\sum_{i=0}^{n'}\sum_{j=0}^{m'}\frac{
    {n \choose n'-i}{m \choose m'-j}}{i!j!}\cos\bigl(\tfrac{\pi}{2}l\bigr)\left[\frac{1}{\eta}\Gammalower(\tfrac{l+1}{2},\eta^2) - \frac{1}{\eta^{3}}\Gammalower(\tfrac{l+3}{2},\eta^2)\right]
\end{equation}

The lowest order contribution in all of these cases is at $\eta_z=0$, where
\[
\begin{aligned}
\bm I^{(xx)}_{l}(0,0,0)&=\frac{4\pi^2 \cos\left(\tfrac{\pi}{2}l\right) (l+2)}{(l+1) (l+3)}\\
\bm I^{(zz)}_{l}(0,0,0)&=\frac{8\pi^2 \cos\left(\tfrac{\pi}{2}l\right)}{(l+1) (l+3)}.
\end{aligned}
\]

Since we reduce to one-dimensional physics in the main text we also define $\eta=\eta_z$ with $\eta_x=\eta_y=0$ from here on out.

\section{Thermal density matrices}
\label{App:thermal-density-matrices}
In this section of the appendix we give some details on the numerical sampling of thermal states by performing Metropolis-Monte-Carlo sampling of the canonical partition function.\\
We consider thermal states with respect to $H_{\text{at}}$, which are mixtures of product states with $\Ne$ excited particles and $\Nm=\Ne+\Ng$ total particles
\begin{equation}
    \begin{aligned}
\rho(T)&=\sum_{\substack{n_0,n_1,\ldots |\sum n_i=\Ng\\m_0,m_1,\ldots |\sum m_i=\Ne}}\me^{-\beta \hbar\omega(\bm n,\bm m)}
\left[\prod_{i=0}^\infty \left(\hat g^\dagger_{i}\right)^{n_i}\right]
\left[\prod_{i=0}^\infty \left(\hat e^\dagger_{i}\right)^{m_i}\right]\outerproduct{0}{0}
\left[\prod_{i=0}^\infty \left(\hat g_{i}\right)^{n_i}\right]
\left[\prod_{i=0}^\infty \left(\hat e_{i}\right)^{m_i}\right]\\
&=\sum_{\substack{n_0,n_1,\ldots |\sum n_i=\Ne\\m_0,m_1,\ldots |\sum m_i=\Ng}}\me^{-\beta\hbar \omega(\bm n,\bm m)}\outerproduct{\psi(\bm n,\bm m)}{\psi(\bm n,\bm m)},
\end{aligned}
\end{equation}
with the inverse temperature $\beta=1/k_BT$ and the frequency
\[
\omega(\bm n,\bm m)=\sum_{j=0}^\infty j\nu(m_j+n_j).
\]

These are two independent canonical ensembles for the ground and excited type particles with fixed particle numbers. Note that we do not include the electronic energy $\omega_0$ since we do not assume that the ratio between ground and excited state particles is thermal, but can be set independently of temperature. 

Since it is impossible to obtain exact expressions for the thermal density matrices, we perform a Metropolis Monte-Carlo sampling of the canonical partition function.
\[
\rho(T)=\frac{1}{N_s}\sum_{j=1}^{N_s}\outerproduct{\psi^{(j)}}{\psi^{(j)}},
\]
with $N_s$ samples. 

\section{Efficient calculation of $p_{S,M}$ with spin addition for the tight-trap regime}
\label{App:tight-trapd}

Recalling the definition of $p_{S,M}$, we define the decomposition for the thermal and Monte-Carlo terms $p_{S,M}(\bm n,\bm m)$ and $p_{S,M}^{(j)}$. Then we can define thermal populations

\[
p_{S,M}(T)=\sum_{\substack{n_0,n_1,\ldots  |\sum n_i=\Ng\\m_0,m_1,\ldots  |\sum m_i=\Ne}}\me^{-\beta \omega(\bm n,\bm m)}p_{S,M}(\bm n,\bm m)=\lim_{N_s\to\infty}\frac{1}{N_s}\sum_{j=1}^{N_s}p_{S,M}^{(j)}.
\]

Using these thermal states as initial states for the time evolution of the ensemble, i.e., defining $p_{S,M}(T,t=0)=p_{S,M}(T)$, its time evolution is due to the superradiant master equation

\[
\dot p_{S,M}(t;T)=-\gamma h_{S,M}p_{S,M}(t;T)+\gamma h_{S,M+1}p_{S,M+1}(t;T).
\]

Thermal collective observables, such as the intensity, are then found as

\[
I(t;T)=\sum_{S,M}\gamma h_{S,M}p_{S,M}(t;T).
\]

\section{Matrix elements in the Lamb-Dicke regime}
\label{App:LambDicke}

If we expand the Franck-Condon factors in a Taylor series, we obtain to second order
\[
\eta_{n,m}=\delta_{n,m}+\iu \eta \left( \sqrt{m}    \delta_{n,m-1} + \sqrt{m+1}\delta_{n,m+1} \right)
 - \frac{\eta^2}{2} \big[(2n+1) \delta_{n,m} + \sqrt{(m+1)(m+2)}\delta_{n,m+2}
+\sqrt{m(m-1)}\delta_{n,m-2}\big)\big]+ \mathcal{O}(\eta^3),
\]
we obtain equivalent results from the decay rate matrix as derived in general in Sec.\ref{App:matrix_eta}, expanded to second order in $\eta$. Indeed, to second order in $\eta$ the nonzero matrix elements read

\[
\begin{aligned}
    \Gamma_{n,n}^{m,m}&=\gamma\left(1-\frac{2}{5}\eta^2(1+m+n)\right)+\mathcal O\left(\eta^4\right)\\
    \Gamma_{n+1,n}^{m+1,m}&=\gamma\frac{2}{5}\eta^2\sqrt{(1+m)(1+n)}\\
    \Gamma_{n+1,n}^{m-1,m}&=\gamma\frac{2}{5}\eta^2\sqrt{m(1+n)}\\
    \Gamma_{n+2,n}^{m,m}&=\gamma\frac{1}{5}\eta^2\sqrt{m(m-1)},
\end{aligned}
\]
where the other matrix elements are obtained by considering the adjoints.

We thus define collective operators as in the main text
\[
\begin{aligned}
    \Sig{0}&=\sum_{n=0}^\infty\gd{n}\e{n}\quad&&\Sig{-}=\sum_{n=1}^\infty \sqrt{n}\gd{n-1}\e{n}\\
    \Sig{+}&=\sum_{n=0}^\infty \sqrt{n+1}\gd{n+1}\e{n}\quad&&\Sig{\text{lin}}=\sum_{n=0}^\infty n\gd{n}\e{n}\\
    \Sig{++}&=\sum_{n=0}^\infty \sqrt{(n+1)(n+2)}\gd{n+2}\e{n}\quad&&\Sig{--}=\sum_{n=2}^\infty \sqrt{n(n-1)}\gd{n-2}\e{n},
\end{aligned}
\]
leading to dissipative dynamics with
\[
\begin{aligned}
\mathcal L[\rho]&=\sum_{\alpha,\beta\in\{0,+,-,\text{lin},++,--\}}\Gamma_{\alpha,\beta}\left(\Sig{\alpha}\rho\Sigd{\beta}-\frac{1}{2}\!\anticommutator{\Sigd{\beta}\Sig{\alpha}}{\rho}\right),
\end{aligned}
\]
where the matrix elements of the $\Gamma^{(\text{LD})}$-matrix then become explicitly, using $\Gamma_0^{(\text{LD})}=\gamma(1 - \frac{2\eta^2}{5})$ and $\Gamma_1^{(\text{LD})}=\frac{2}{5}\eta^2\gamma_0$

\[
\bm\Gamma^{(\text{LD})}= 
\begin{pmatrix}
    \Gamma_0^{(\text{LD})} & 0 & 0 & -\Gamma_1^{(\text{LD})} & \Gamma_1^{(\text{LD})} / 2 & \Gamma_1^{(\text{LD})}/2 \\
    0 & \Gamma_1^{(\text{LD})} & \Gamma_1^{(\text{LD})} & 0 & 0 & 0\\
    0 & \Gamma_1^{(\text{LD})} & \Gamma_1^{(\text{LD})} & 0 & 0 & 0\\
    -\Gamma_1^{(\text{LD})} & 0 & 0 & 0 & 0 & 0\\ 
    \Gamma_1^{(\text{LD})}/2 & 0 & 0 & 0 & 0 & 0\\ 
    \Gamma_1^{(\text{LD})}/2 & 0 & 0 & 0 & 0 & 0\\ 
\end{pmatrix}
\]
One has to be careful when diagonalizing this matrix, since including all orders in $\eta$ in the diagonalization leads to spurious orders when multiplying the diagonalized decay channels back into the original ones. At second order in $\eta$ the couplings to $\Sig{\text{lin}}$, $\Sig{\text{++}}$ and $\Sig{\text{lin}}$ are purely offdiagonal and would induce spurious $\eta^4$ terms in the diagonalized picture. They must thus be omitted for the sake of consistency so that we may assume a dissipative coupling matrix
\[
\bm\Gamma=
\begin{pmatrix}
    \Gamma_0^{(\text{LD})} & -\Gamma_1^{(\text{LD})} & 0 & 0\\
    -\Gamma_1^{(\text{LD})} & 0 & 0 & 0 \\
    0 & 0 & \Gamma_1^{(\text{LD})} & \Gamma_1^{(\text{LD})} \\
    0 & 0 &\Gamma_1^{(\text{LD})} & \Gamma_1^{(\text{LD})}
\end{pmatrix},
\]
as in the main text which only couples $\{0,+,-\}$. Diagonalizing this matrix, keeping only terms up to $\eta^2_z$ gives the following diagonal decay channels with the associated rates absorbed into the collapse operator
\[
\begin{aligned}
    \hat{L}_1 &= \sqrt{\gamma}\sum_{n=0}^{\infty}\sqrt{1-\frac{2}{5}\eta^2(2n + 1)}\gd{n}\e{n}\\
\hat{L}_2 &= \sqrt{\gamma\frac{2}{5}\eta^2}\left(\Sig{+} + \Sig{-}\right)\\
    \hat{L}_3 &=  0 \cdot \frac{1}{\sqrt{2}}\left(\Sig{+} - \Sig{-}\right)
\end{aligned}
\]
The collective collapse operator $\hat{L}_3$ has a decay rate of $0$ and is therefore dark.

\section{Matrix elements beyond the Lamb-Dicke regime}\label{App:beyond_instantaneous}

\emph{Matrix elements for bosons: }We can give simplified expressions for the matrix elements beyond the Lamb-Dicke regime, under the $T=0$ restriction, and for fermions, further restrict to $\Ne \leq \Ng$

For bosons at $T=0$ we are only interested in matrix elements of the form
\[
\eta_{0,n}(k)=\matrixelement{0}{\me^{-\iu k \hat z}}{n}=\matrixelement{0}{\mathcal D[-\iu k z_{\rm{zpm}}/\sqrt{2}]}{n} = \me^{-\frac{\eta^2}{2}}\frac{\eta^n}{\sqrt{n!}},
\]
so that the matrix elements become a simple power series in $\eta$
\[
\Gamma_{m,0}^{n,0}(\eta)
= \gamma\eta^{\,m+n}\cos\bigl(\tfrac{\pi}{2}(m+n)\bigr)\frac{(-1)^n}{\sqrt{m!n!}}\frac{3}{2}\sum_{j=0}^{\infty}
\left[\frac{(-1)^j}{j!}\frac{m+n+2j+2}{(m+n+2j+1)(m+n+2j+3)}\right] \eta^{2j}.
\]

Notably, for odd powers of $n+m$ this is zero, making it a power series in $\eta^2$. If we are only interested in the instantaneous intensity, the $n=m$ restriction yields the simplification
\[
\Gamma_{n,0}^{n,0}(\eta)
= \gamma\frac{\eta^{2n}}{n!}\frac{3}{2}\sum_{j=0}^{\infty}
\left[\frac{(-1)^j}{j!}\frac{2n+2j+2}{(2n+2j+1)(2n+2j+3)}\right] \eta^{2j}.
\]

For "on-site" decay rate for the ground state of the trap is given by
\[
\Gamma_{0,0}^{0,0}(\eta)
= \gamma \frac{3}{2}\sum_{j=0}^{\infty}
\left[\frac{(-1)^j}{j!}\frac{2j+2}{(2j+1)(2j+3)}\right] \eta^{2j} = \gamma \left(\frac{3\sqrt{\pi}(\eta^2 + 1/2)\erf(\eta)}{8\eta^3} - \frac{3\me^{-\eta^2}}{8\eta^2}\right).
\]
Note that for $\eta \rightarrow 0$ only the $j=0$ term survives in the sum, which gives $2/3$, canceling the $3/2$ prefactor, and the decay rate becomes simply the usual free space rate $\gamma$. This serves nicely as a sanity check, as for $\eta\rightarrow 0$ the particle becomes delta-localized in the trap, which should be equivalent to the usual treatment of spontaneous emission of a particle at some fixed position. Further, for $\eta\rightarrow\infty$, the matrix element decays as $\eta^{-1}$.\\
Finally, we want to know the total decay rate out of the ground state of the trap, which is given by 
\[
\begin{aligned}
\sum_{n=1}^{\infty}\Gamma_{n,0}^{n,0} &= \gamma\sum_{n=1}^{\infty}\frac{\eta^{2n}}{n!}\frac{3}{2}\sum_{j=0}^{\infty}
\left[\frac{(-1)^j}{j!}\frac{2n+2j+2}{(2n+2j+1)(2n+2j+3)}\right] \eta^{2j}\\
&= \gamma\left[1 - \frac{3\sqrt{\pi}}{8\eta^3}\left(\eta^2 + \frac{1}{2}\right)\erf(\eta) + \frac{3\me^{-\eta^2}}{8\eta^2}\right].
\end{aligned}
\]
As a sanity check, we obtain for $\eta \rightarrow 0$ (using $\erf(\eta) \approx 2\eta/\sqrt{\pi}$), the total decay rate out of the level $\sum_{n=1}^{\infty} \Gamma_{n,0}^{n,0} =0$. Further, for $\eta\rightarrow \infty$ we obtain $\gamma$, as all terms decay with $\eta^{-\alpha}$ with $\alpha\geq1$.

\emph{Matrix elements for fermions: }For fermions, we require the general matrix elements (assuming $n_z\geq n_z'$ and $m_z\geq m_z'$)
\[
\begin{aligned}
    \bm \Gamma^{m_z,m'_z}_{n_z,n'_z}&=(-1)^{n_z-n_z'}\gamma \frac{3}{8\pi^2}\sqrt{\frac{n_z'!m_z'!}{n_z!m_z!}}\iu^{n_z+m_z-n_z'-m_z'}\\
    &\times \sum_{i=0}^{n_z'}\sum_{j=0}^{m_z'}
    {n_z \choose n_z'-i}{m_z \choose m_z'-j}\frac{\eta^{n+m-n'-m'+2(i+j)}}{i!j!}\bm I_{n_z+m_z-n_z'-m_z'+2(i+j)}(\eta_x,\eta_y,\eta).
\end{aligned}
\]

We consider, for simplicity, $\bm I_{l}^{(zz)}(0,0,\eta)$ 
\[
    \bm I^{(zz)}_{l}(0,0,\eta) =-8\iu\pi^2 \cos\left(\tfrac{\pi}{2}l\right)\sum_{p=0}^\infty \frac{(-1)^p}{p!}\frac{\eta^{2p}}{(l+2p+1)(l+2p+3)}.
\]
Restricting again to the diagonal part $m_z' = n_z', m_z = n_z$ and using $l = 2(n-m + i+ j)$ with $n\geq m$ we obtain
\[
\Gamma^{n,m}_{n,m} =\gamma \frac{3}{8\pi^2}\frac{m!}{n!} \sum_{i=0}^{m}\sum_{j=0}^{m}\frac{
    {n \choose m-i}{n \choose m-j}}{i!j!}\cos\bigl(\tfrac{\pi}{2}l\bigr)\left[\frac{1}{\eta}\Gammalower(\tfrac{l+1}{2},\eta^2) - \frac{1}{\eta^{3}}\Gammalower(\tfrac{l+3}{2},\eta^2)\right]
\]

\section{Symmetry-guided perturbative basis expansion}\label{app:perturbative_basis}

Consider first a fairly general case to illustrate the counting procedure for perturbative orders when doing Monte Carlo wave function dynamics. We have purely dissipative dynamics with two dissipators $\hat L_0, \eta \hat L_1$ where we aim to determine the time evolution up to some order in $\eta$. The nonhermitian Hamiltonian is then $\mathcal H_{\text{nh}}=-\frac{\iu}{2}\left(\hat L_0^\dagger \hat L_0+\eta^2\hat L_1^\dagger \hat L_1\right)$

\begin{itemize}
    \item Each action of $\hat L_1^\dagger \hat L_1$ in the nonhermtian Hamiltonian yields a factor $\eta^2$, increasing the order by two.
    \item Each quantum jump with $\hat L_1$ also provides a contribution of order $\eta^2$ since the jump probability scales with $\eta^2$. Thus, this also raises the order by two.
\end{itemize}

As such, a perturbative basis can now be generated.

\begin{enumerate}
    \item Start with initial state $\mathcal B=\{\psi_0\}$.
    \item Form a Basis $\mathcal B'=\text{span}\left\{\hat L_0^n b,\left(\hat L_0^\dagger\hat L_0\right)^nb\,\,\middle\vert\,\, \forall b\in \mathcal B,n=0,1,\ldots\right\}$. I.e. find the invariant subspace under the action of the unperturbed time evolution. Note that this does not contribute any order of $\eta$.
    \item Find $\mathcal B_j=\hat L_1 \mathcal B'$ and $\mathcal B_h=\hat L_1^\dagger \hat L_1 \mathcal B'$. Their union is the update for $\mathcal B=\mathcal B'\oplus \mathcal B_j\oplus \mathcal B_h$.
    \item Repeat steps (2) and (3) with a penalty of $\eta^2$ for each repetition until the desired order is reached.
\end{enumerate}

For the more general case when we have mixed order jump terms such as $\eta^n \hat L_n$ the counting is adjusted in the obvious way, penalizing the action with this operator in the jump or nonhermtian Hamiltonian by $\eta^{2n}$. Combinatorially, one must then find the different action paths that yield the desired order.

Consider the example $\{\hat L_0,\eta \hat L_1,\eta^2 \hat L_2\}$ then define the basis update operation as $\mathcal U(\hat L_i)$ for a single action and $\mathcal U_c(\hat L_i)$ for arbitrarily many actions. Then to order six, the different paths required are

\[
\begin{aligned}
    \mathcal B=
    \mathcal U_c(\hat L_0)\Big[&\mathcal U(\hat L_1)\mathcal U_c(\hat L_0)\mathcal U(\hat L_1)\mathcal U_c(\hat L_0)\mathcal U(\hat L_1)\\
    &\oplus\mathcal U(\hat L_2)\mathcal U_c(\hat L_0)\mathcal U(\hat L_1)\\
    &\oplus\mathcal U(\hat L_1)\mathcal U_c(\hat L_0)\mathcal U(\hat L_2) \Big]\mathcal U_c(\hat L_0)\{\psi_0\}.
\end{aligned}
\]

The procedure constructs the perturbatively reachable subspace in a matrix-free way. Only after this basis is fixed, do we assemble the Hamiltonian and jump operators as sparse matrices and perform Krylov time evolution. The reduction comes from the fact that closure under $L_0$ is often small compared to the full Hilbert space, so higher-order $\eta$ contributions only add a limited number of new directions.

\section{Thermodynamic limit for instantaneous intensity for fermions}
\label{App:instantaneous_intensity_fermions}
For fermionic systems at zero temperature the instantaneous intensity can be rewritten as
\[
\begin{aligned}
\gamma_\text{F}(\eta,\Ng)&=\frac{3\gamma}{2\Ne\eta}\int_{0}^{\eta}\diff{u}(1-u^2/\eta^2)\sum_{m=0}^{\Ne}\sum_{n=0}^{\Ng}\abs{\matrixelement{m}{\me^{\iu u(\hat a-\hat a^\dagger)}}{n}}^2
\end{aligned}
\]
which puts our intuition into a transparent formula. What obstructs fermionic emission are overlaps of a displacement between occupied levels, i.e. motional processes blocked by the exclusion rule. What this also makes clear is the asymptotic behavior $\eta\to\infty$ at fixed $\Ne,\Ng$ corresponding to the zero density limit. There the integration boundary can be replaced by $\infty$ and 

\[
\begin{aligned}
\gamma_\text{F}(\eta,\Ng)&=\frac{C_\text{F}(\Ng)}{\Ne\eta}+\mathcal O(\eta^{-2})\\
C_\text{F}(\Ng)&=\frac{3\gamma}{2}\int_{0}^{\infty}\diff{u}\sum_{m=0}^{\Ne}\sum_{n=0}^{\Ng}\abs{\matrixelement{m}{\me^{\iu u(\hat a+\hat a^\dagger)}}{n}}^2.
\end{aligned}
\]

As opposed to bosonic systems there is no linear scaling in the density. Indeed, keeping one $\Ne$ or $\Ng$ yields a blocking factor that becomes constant when increasing the particle number since both the excited or ground state emitters do not see the other fermions. Increasing both $\Ne$ and $\Ng$ at the same time leads to a linear increase in $C_F(\Ng)$ restoring the proportionality to $\Ne$ so that perfect blocking can survive in this limit. Since there is no proportionality to $\Ng$ however, there is no linear dependence on the density. This is however misleading since the order of limiting procedures becomes important here. Keeping constant density implies setting $\eta=\chi \sqrt{\mathcal N}$. and then taking the limit $\mathcal N\to\infty$. Formally then

\[
\begin{aligned}
\gamma_\text{F}^\infty&\sim\lim_{\mathcal N\to\infty}\frac{3\gamma}{2\Ne\chi \sqrt{\Nm}}\int_{0}^{\chi\sqrt{\mathcal N}}\diff{u}\sum_{m=0}^{\Ne}\sum_{n=0}^{\Ng}\abs{\matrixelement{m}{\me^{\iu u(\hat a+\hat a^\dagger)}}{n}}^2\,.
\end{aligned}
\]

We shall assume that both $\Ne$ and $\Ng$ scale linearly with $\Nm$, i.e. a finite excitation fraction. This form motiatates the arguments presented in the main text in terms of momentum quantum numbers. Geometrically, the summation constitutes the overlap of a disk with radius $\sqrt{2\Ne}$ with a disk with radius $\sqrt{2\Ng}$, but displaced by $u$. We replace $\Ng$ and $\Ne$ by $\Nm$ for this scaling argument and use the geometric overlap of two circles of radius $R$ with distance $u$ as $A_R(u)=2R^2 \arccos\left(\frac{u}{2R}\right) - \frac{u}{2} \sqrt{4R^2 - u^2}$ so that 

\[
\gamma_\text{F}^\infty\sim\lim_{\mathcal N\to\infty}\frac{3\gamma}{2\Ne\chi \sqrt{\Nm}}\int_{0}^{\chi\sqrt{\mathcal N}}\diff{u}A_{\sqrt{2N}}(u)\,.
\]

Making the substitution $v=u/\sqrt{\Nm}$ we obtain
\[
\gamma_\text{F}^\infty\sim\frac{3\gamma}{2\chi}\int_{0}^{\chi}\diff{v}A_{\sqrt{2}}(v)\,.
\]
which is independent of $\Nm$ and hence finite.

\section{Slope of the intensity at time zero}
\label{App:beyond_intensity}

In this section, we give some details on the calculation of the initial slope of the intensity for arbitrary trap-width.\\
To certify whether the intensity increases for small times, we determine the time evolution of the intensity operator 
\[
\hat I=\sum_{n,m,n',m'}\Gamma_{n,n'}^{m,m'}\Sigd{n,n'}\Sig{m,m'}
\]
according to the equations of motion
\[\begin{aligned}
\dot{\expval{I}}&=\frac{1}{2}\sum_{\substack{n_1,m_1,n'_1,m'_1\\n_2,m_2,n'_2,m'_2}}\Gamma_{n_1,n'_1}^{m_1,m'_1}\Gamma_{n_2,n_2'}^{m_2,m_2'}
\expval{\Sigma_{n_2,n_2'}^\dagger\commutator{\Sigd{n_1,n'_1}\Sig{m_1,m'_1}}{\Sigma_{m_2,m_2'}}
+\commutator{\Sigma_{n_2,n_2'}^\dagger}{\Sigd{n_1,n'_1}\Sig{m_1,m'_1}}\Sigma_{m_2,m_2'}}\\    
&=\frac{1}{2}\sum_{\substack{n_1,m_1,n'_1,m'_1\\n_2,m_2,n'_2,m'_2}}\Gamma_{n_1,n'_1}^{m_1,m'_1}\Gamma_{n_2,n_2'}^{m_2,m_2'}
\left[\expval{\ed{n_2'}\g{n_2}\commutator{\ed{n'_1}\g{n_1}\gd{m_1}\e{m'_1}}{\gd{m_2}\e{m_2'}}}
+\expval{\commutator{\ed{n_2'}\g{n_2}}{\ed{n'_1}\g{n_1}\gd{m_1}\e{m'_1}}\gd{m_2}\e{m_2'}}\right]
\end{aligned}
\]

The commutators can be performed 
\[
\begin{aligned}
    \commutator{\ed{n'_1}\g{n_1}\gd{m_1}\e{m'_1}}{\gd{m_2}\e{m_2'}}=&\gd{m_2}\g{n_1}\gd{m_1}\commutator{\ed{n'_1}\e{m'_1}}{\e{m_2'}}+\ed{n'_1}\e{m'_1}\commutator{\g{n_1}\gd{m_1}}{\gd{m_2}}\e{m_2'}\\
    &=\ed{n'_1}\e{m'_1}\gd{m_1}\e{m_2'}\delta_{n_1,m_2}-\gd{m_2}\g{n_1}\gd{m_1}\e{m'_1}\delta_{n_1',m_2'}
\end{aligned}
\]
and
\[
\begin{aligned}
\commutator{\ed{n_2'}\g{n_2}}{\ed{n'_1}\g{n_1}\gd{m_1}\e{m'_1}}&=\ed{m_2'}\commutator{\g{m_2}}{\g{n_1}\gd{m_1}}\ed{n'_1}\e{m'_1}+\commutator{\ed{n_2'}}{\ed{n'_1}\e{m'_1}}\g{n_1}\gd{m_1}\g{n_2}\\
&=\ed{n_2'}\g{n_1}\ed{n'_1}\e{m'_1}\delta_{n_2,m_1}-\ed{n'_1}\g{n_1}\gd{m_1}\g{n_2}\delta_{n_2',m'_1}
\end{aligned}
\]

Plugging them back into the derivative of the intensity
\[\begin{aligned}
\dot{\expval{I}}&=\frac{1}{2}\sum_{\substack{n_1,m_1,n'_1,m'_1\\n_2,n'_2,m'_2}}\Gamma_{n_1,n'_1}^{m_1,m'_1}\Gamma_{n_2,n_2'}^{n_1,m_2'}
\expval{\ed{n_2'}\ed{n'_1}\e{m'_1}\e{m_2'}}\expval{\g{n_2}\gd{m_1}}\\
&-\frac{1}{2}\sum_{\substack{n_1,m_1,n'_1,m'_1\\n_2,m_2,n'_2}}\Gamma_{n_1,n'_1}^{m_1,m'_1}\Gamma_{n_2,n_2'}^{m_2,n_1'}
\expval{\g{n_2}\gd{m_2}\g{n_1}\gd{m_1}}\expval{\ed{n_2'}\e{m'_1}}\\
&+\frac{1}{2}\sum_{\substack{n_1,m_1,n'_1,m'_1\\n_2,n'_2,m'_2}}\Gamma_{n_1,n'_1}^{m_1,m'_1}\Gamma_{n_2,n_2'}^{m_1,m_2'}
\expval{\ed{m_2'}\ed{n'_1}\e{m'_1}\e{n_2'}}\expval{\g{n_1}\gd{n_2}}\\
&-\frac{1}{2}\sum_{\substack{n_1,m_1,n'_1,m'_1\\n_2,m_2,n'_2}}\Gamma_{n_1,n'_1}^{m_1,m'_1}\Gamma_{n_2,n_2'}^{m_2,m_1'}
\expval{\g{n_1}\gd{m_1}\g{m_2}\gd{n_2}}\expval{\ed{n'_1}\e{n_2'}}
\end{aligned}\]
Since we consider thermal initial states constructed from mixtures of pure states, we know that each level is always occupied by a Fock state. Thus, the correlation functions can now be evaluated under the assumption of some fixed Fock state occupation per level since, for $i\neq j$
\[
\begin{aligned}
    \expval{\ed{i}\ed{j}\e{i}\e{j}}_B&=\Ne_i\Ne_j\\
    \expval{\ed{i}\ed{j}\e{j}\e{i}}_B&=\Ne_i\Ne_j\\
    \expval{\ed{j}\ed{j}\e{j}\e{j}}_B&=\Ne_j(\Ne_j-1)\\
    \expval{\ed{i}\ed{j}\e{i}\e{j}}_F&=-\Ne_i\Ne_j\\
    \expval{\ed{i}\ed{j}\e{j}\e{i}}_F&=\Ne_i\Ne_j\\
    \expval{\ed{j}\ed{j}\e{j}\e{j}}_F&=0\\
    \expval{\g{i}\gd{i}\g{j}\gd{j}}_B&=(\Ng_j+1)(\Ng_i+1)\\
    \expval{\g{i}\gd{j}\g{j}\gd{i}}_B&=\Ng_j(\Ng_i+1)\\
    \expval{\g{j}\gd{j}\g{j}\gd{j}}_B&=(\Ng_j+1)^2\\
    \expval{\g{i}\gd{i}\g{j}\gd{j}}_F&=(\Ng_j-1)(\Ng_i-1)\\
    \expval{\g{i}\gd{j}\g{j}\gd{i}}_F&=(1-\Ng_i)\Ng_j\\
    \expval{\g{j}\gd{j}\g{j}\gd{j}}_F&=(\Ng_j-1)^2
\end{aligned}
\]
so that
\[
\begin{aligned}
    \expval{\ed{n_2'}\ed{n_1'}\e{m_1'}\e{m_2'}}_B&=(\delta_{n_2',m_1'}\delta_{n_1',m_2'}+\delta_{n_2',m_2'}\delta_{n_1',m_1'})\Ne_{m_1'}\Ne_{n_2'}-\delta_{n_2',m_1'}\delta_{n_1',m_2'}\delta_{n_2',m_2'}\Ne_{n_2'}(\Ne_{n_2'}+1)\\
    \expval{\ed{n_2'}\ed{n_1'}\e{m_1'}\e{m_2'}}_F&=(\delta_{n_2',m_2'}\delta_{n_1',m_1'}-\delta_{n_2',m_1'}\delta_{n_1',m_2'})\Ne_{m_1'}\Ne_{n_2'}\\
    \expval{\g{n_2}\gd{m_2}\g{n_1}\gd{m_1}}_B&=(\Ng_{n_2}+1)(\Ng_{n_1}+1)\,\delta_{n_2 m_2}\delta_{n_1 m_1}+\Ng_{n_1}(\Ng_{n_2}+1)\delta_{n_2 m_1}\delta_{n_1 m_2}-\Ng_{m_2}(\Ng_{m_2}+1)\delta_{m_2 m_1}\delta_{m_2 n_2}\delta_{m_2 n_1}\\
    \expval{\g{n_2}\gd{m_2}\g{n_1}\gd{m_1}}_F&=(1-\Ng_{n_2})(1-\Ng_{n_1})\delta_{n_2 m_2}\delta_{n_1 m_1}+ \Ng_{n_1}(1-\Ng_{n_2})\delta_{n_2 m_1}\delta_{n_1 m_2}
\end{aligned}
\]
which may be combined into the relations
\[
\begin{aligned}
    \expval{\ed{n_2'}\ed{n_1'}\e{m_1'}\e{m_2'}}&=(\delta_{n_2',m_2'}\delta_{n_1',m_1'}+\zeta\delta_{n_2',m_1'}\delta_{n_1',m_2'})\Ne_{m_1'}\Ne_{n_2'}-\frac{1+\zeta}{2}\delta_{n_2',m_1'}\delta_{n_1',m_2'}\delta_{n_2',m_2'}\Ne_{n_2'}(\Ne_{n_2'}+1)\\
    \expval{\g{n_2}\gd{m_2}\g{n_1}\gd{m_1}}&=(1+\zeta\Ng_{n_2})(1+\zeta\Ng_{n_1})\delta_{n_2 m_2}\delta_{n_1 m_1}+ \Ng_{n_1}(1+\zeta\Ng_{n_2})\delta_{n_2 m_1}\delta_{n_1 m_2}\\
    &-\frac{1+\zeta}{2}\Ng_{m_2}(\Ng_{m_2}+1)\delta_{m_2 m_1}\delta_{m_2 n_2}\delta_{m_2 n_1}
\end{aligned}
\]

and using the same indices for all summations leads to (using $\chi=\frac{1+\zeta}{2}$)

\[
\begin{aligned}
  \dot{\expval{I}}&=\sum _{i_1,i_2,i_3} -\chi  \Ne_{i_3} \left(\left(\Gamma _{i_2,i_3}^{i_1,i_3}\right)^2
   \left(\Ne_{i_3}+1\right) \left(\zeta  \Ng_{i_2}+1\right)\right.\\
   &+\left.\Gamma _{i_1,i_3}^{i_2,i_3} \Gamma _{i_2,i_3}^{i_1,i_3}
   \left(\Ne_{i_3}+1\right) \left(\zeta  \Ng_{i_2}+1\right)-\Gamma _{i_2,i_3}^{i_2,i_1} \left(\Gamma _{i_2,i_1}^{i_2,i_3}+\Gamma
   _{i_2,i_3}^{i_2,i_1}\right) \Ng_{i_2} \left(\Ng_{i_2}+1\right)\right)\\
   &+\sum _{i_1,i_2,i_3,i_4}  \Ne_{i_4} \left(\zeta  \Gamma _{i_2,i_3}^{i_1,i_4} \left(\Gamma _{i_1,i_4}^{i_2,i_3}+\Gamma
   _{i_2,i_3}^{i_1,i_4}\right) \Ne_{i_3} \left(\zeta  \Ng_{i_2}+1\right)\right.\\
   &\left.-\Ng_{i_3} \left(-\zeta  \left(\Gamma_{i_1,i_2}^{i_3,i_2}+\Gamma _{i_3,i_2}^{i_1,i_2}\right) \Gamma _{i_3,i_4}^{i_1,i_4} \Ne_{i_2}+\left(\Gamma _{i_2,i_4}^{i_3,i_1}\right)^2
   \left(\zeta  \Ng_{i_2}+1\right)+\Gamma _{i_3,i_1}^{i_2,i_4} \Gamma _{i_2,i_4}^{i_3,i_1} \left(\zeta  \Ng_{i_2}+1\right)\right.\right.\\
   &\left.\left.+\zeta \left(\Gamma _{i_1,i_2}^{i_1,i_4}+\Gamma _{i_1,i_4}^{i_1,i_2}\right) \Gamma _{i_3,i_4}^{i_3,i_2} \left(\zeta  \Ng_{i_1}+1\right)\right)
   +\Gamma_{i_1,i_2}^{i_3,i_2} \Gamma _{i_3,i_4}^{i_1,i_4} \Ne_{i_2}\right.\\
   &+\Gamma _{i_3,i_2}^{i_1,i_2} \Gamma _{i_3,i_4}^{i_1,i_4}\left.\Ne_{i_2}-\zeta  \Gamma _{i_1,i_2}^{i_1,i_4} \Gamma _{i_3,i_4}^{i_3,i_2} \Ng_{i_1}-\zeta  \Gamma _{i_1,i_4}^{i_1,i_2} \Gamma
   _{i_3,i_4}^{i_3,i_2} \Ng_{i_1}-\Gamma _{i_1,i_2}^{i_1,i_4} \Gamma _{i_3,i_4}^{i_3,i_2}-\Gamma _{i_1,i_4}^{i_1,i_2} \Gamma_{i_3,i_4}^{i_3,i_2}\right)  
\end{aligned}
\]

To consider this sum numerically and analytically, it is useful to point out some structural points. We first define a tensor
\[
\begin{aligned}
C^{m_1,m_2,m_1',m_2'}_{n_1,n_2,n_1',n_2'}
[f_1(k,k'),f_2(k,k'),f_3(k,k'),f_4(k,k')]
&=\mathcal C
\int \diff{\bm k}\diff{\bm k'}\,
\matrixelement{m_1}{\exp(\iu f_1\hat z)}{n_1}
\matrixelement{m_2}{\exp(\iu f_2\hat z)}{n_2}
\\
&\qquad\qquad\times
\matrixelement{m_1'}{\exp(\iu f_3\hat z)}{n_1'}
\matrixelement{m_2'}{\exp(\iu f_4\hat z)}{n_2'}
\,
\widetilde{\mathbf G}(\bm k)\,
\widetilde{\mathbf G}(\bm k')
\end{aligned}
\]
where each $f_i=a_i k+b_i k'$ real functions with coefficients $a_i,b_i \in\{0,\pm 1\}$ and we realize that
\[
\Gamma_{n_1,n_2}^{m_1,m_2}\Gamma_{n_1',n_2'}^{m_1',m_2'}
=C_{n_2,m_1,n_2',m_1'}^{n_1,m_2,n_1',m_2'}[-k,k,-k',k']
\]

We can now perform contractions like
\[
\sum_r C_{m_1,r,m_1',m_2'}^{n_1,n_2,r,n_2'}[f_1,f_2,f_3,f_4]=C_{m_1,m_1',m_2'}^{n_1,n_2,n_2'}[f_1,f_2+f_3,f_4],
\]
adding momenta of the contracted indices and pairing the leftover indices from the contraction. Leaving out indices and momenta means that they do not appear in the integration. Keeping in mind that
\[
C_{m_1,m_2,m_1',m_2'}^{n_1,n_2,n_1',n_2'}[f_1,f_2,f_3,f_4]
=C_{n_1,m_2,m_1',m_2'}^{m_1,n_2,n_1',n_2'}[-f_1,f_2,f_3,f_4]
\]
indices at the same position can be swapped at the expense of a minus sign via complex conjugation of the relevant matrix element. The same holds for indices and momenta in the other slots.

Thus, contracting where possible ultimately leads to
\begin{equation}
    \begin{split}
    2\dot I&=-2 \sum _{m_1=0}^{\infty } \gamma ^2 \Ne_{m_1}\\
    &-\chi\sum _{m_1=0}^{\infty }\left(\left(\Ne_{m_1}\right)^2[C_{m_1m_1}^{m_1m_1}(k',-k')+C_{m_1m_1}^{m_1m_1}(k'-k,2 k)]+\Ne_{m_1}[C_{m_1m_1}^{m_1m_1}(k',-k')+C_{m_1m_1}^{m_1m_1}(k'-k,2 k)]\right)\\
    &-2 \sum_{m_1,m_2} \zeta  \Ne_{m_2} \Ng_{m_1} C_{m_2m_1}^{m_1m_2}(-k',k')\\
    &+\sum_{m_1,m_2=0}^{\infty} \left(-\zeta  \chi  \Ng_{m_1} \left(\Ne_{m_2}\right)^2C_{m_2m_1m_2}^{m_2m_1m_2}(k'-k,k,-k')-\zeta\chi  \Ng_{m_1} \left(\Ne_{m_2}\right)^2C_{m_2m_2m_2}^{m_1m_1m_2}(-k,k'-k,-k')\right.\\
    &+\Ne_{m_1}\Ne_{m_2}C_{m_1m_2}^{m_1m_2}(k',-k') - \zeta\Ng_{m_1}\Ne_{m_2}C_{m_1m_2}^{m_1m_2}(k'-k,k)\\
    &+\Ne_{m_1}\Ne_{m_2}C_{m_1m_2}^{m_1m_2}(k'-k,2 k)  +  \zeta\Ne_{m_1}\Ne_{m_2}C_{m_2m_1}^{m_1m_2}(k',-k')\\
    &+\zeta\Ne_{m_1}\Ne_{m_2}C_{m_2m_1}^{m_1m_2}(k'-k,2 k) - \Ng_{m_1}\Ne_{m_2}C_{m_2m_1}^{m_1m_2}(k'-k,k-k')\\
    &-\Ng_{m_1}\Ne_{m_2}C_{m_2m_1}^{m_2m_1}(k-k',k'-k)  -  \zeta\Ng_{m_1}\Ne_{m_2}C_{m_2m_2}^{m_1m_1}(-k,k'-k)\\
    &+\chi\left(\Ng_{m_1}\right)^2\Ne_{m_2}C_{m_2m_1m_1}^{m_1m_1m_2}(-k,k-k',k')  +  \chi\Ng_{m_1}\Ne_{m_2}C_{m_2m_1m_1}^{m_1m_1m_2}(-k,k-k',k')\\
    &-\zeta\chi\Ng_{m_1}\Ne_{m_2}C_{m_2m_1m_2}^{m_2m_1m_2}(k'-k,k,-k')  +  \chi\left(\Ng_{m_1}\right)^2\Ne_{m_2}C_{m_2m_2m_1}^{m_1m_1m_1}(-k,k'-k,-k')\\
    &\left.+\chi\Ng_{m_1}\Ne_{m_2}C_{m_2m_2m_1}^{m_1m_1m_1}(-k,k'-k,-k')  -  \zeta\chi\Ng_{m_1}\Ne_{m_2}C_{m_2m_2m_2}^{m_1m_1m_2}(-k,k'-k,-k')\right)\\
    &+\sum _{m_1,m_2,m_3=0}^{\infty}\left(\zeta  \Ne_{m_1}\Ng_{m_2} \Ne_{m_3} C_{m_1m_1m_3}^{m_2m_2m_3}(-k,k'-k,-k')+\zeta \Ne_{m_1}\Ng_{m_2} \Ne_{m_3} C_{m_1m_2m_3}^{m_1m_2m_3}(k'-k,k,-k')\right.\\
    &-\Ng_{m_1} \Ng_{m_2}\Ne_{m_3} C_{m_1m_3m_2}^{m_1m_3m_2}(k'-k,k,-k')+\Ng_{m_1} \Ne_{m_2} \Ne_{m_3} C_{m_2m_3m_2}^{m_1m_1m_3}(-k,k'-k,-k')\\
    & +\Ng_{m_1}\Ne_{m_2} \Ne_{m_3}C_{m_3m_1m_2}^{m_2m_1m_3}(k'-k,k,-k')-\zeta  \Ng_{m_1} \Ng_{m_2}\Ne_{m_3}C_{m_3m_2m_1}^{m_1m_2m_3}(-k,k-k',k')\\
    &\left.-\Ng_{m_1} \Ng_{m_2}\Ne_{m_3}C_{m_3m_3m_2}^{m_1m_1m_2}(-k,k'-k,-k')-\zeta \Ng_{m_1} \Ng_{m_2} \Ne_{m_3} C_{m_3m_3m_2}^{m_1m_2m_1}(-k,k'-k,-k')\right)
    \end{split}
\end{equation}

All of the remaining $C$-tensors do not contain any zeros and hence
\[
\lim_{\eta\to\infty} C_{n_1,\ldots,n_l}^{m_1,\ldots,m_l}[k_1,\ldots,k_l]=0
\]
for $k_1,\ldots,k_l\neq 0$. This is due to the fact that they are finite Gaussian integrations, which necessarily bring a factor of $1/\eta$ at least. Hence,
\[
\lim_{\eta\to\infty} \dot{\expval{I}}=-\gamma^2 \Ne,
\]
which is independent of temperature or species, and decreases. We can thus conclude that no superradiant burst survives for an infinitely wide trap and finite particle number, which is expected as particle statistics effects must vanish as the density approaches zero.

For the initial condition $\Ng_i=0$, the general expression simplifies significantly to the following expression
\[
\begin{aligned}
\dot{\langle I \rangle} = -&\sum _{m_1=0}^{\infty } \gamma ^2 \Ne_{m_1}\\
-\frac{\chi}{2}&\sum_{m_1=0}^{\infty } \left[\left(\Ne_{m_1}\right)^2 \left( C_{m_1m_1}^{m_1m_1}(k'-k,2 k)  +  C_{m_1m_1}^{m_1m_1}(k',-k')\right) +\Ne_{m_1}\left(
   C_{m_1m_1}^{m_1m_1}(k'-k,2 k) +
   C_{m_1m_1}^{m_1m_1}(k',-k')\right)\right]\\
&\frac{1}{2}\sum_{m_1,m_2=0}^{\infty }\Ne_{m_1} \Ne_{m_2} \left[\zeta C_{m_2m_1}^{m_1m_2}(k'-k,2
   k) +\zeta 
   C_{m_2m_1}^{m_1m_2}(k',-k')+ C_{m_1m_2}^{m_1m_2}(k'-k,2 k)+ C_{m_1m_2}^{m_1m_2}(k',-k')\right]
\end{aligned}
\]

Numerical evaluation of the $C-$tensors can be efficiently performed by utilizing Gauss-Legendre quadratures, since the integration is performed over the interval $[-1,1]$. At $T=0$ we can further simplify the general expressions, which we do for bosons and fermions in the following subsections.

\subsection{Bosons at $T=0$}

At $T=0$ for bosons, the general expression above simplifies to

\[
\begin{aligned}
    2\dot{\expval{I}} &= \frac{\Ne}{2} \Big[ 2\Ng(2\Ne - \Ng - 1) C_3(k, k', k'-k)- 2\Ng \big( C_2(k, k'-k) + C_2(k'-k, k'-k) \big) \\
    &\quad + 2(\Ne - 1) C_2(2k, k'-k)+ 2(\Ne - 1 - \Ng) C_2(k', k') - 2\gamma^2 \Big]
\end{aligned}
\]
where we defined $C_2(f_1,f_2) = C_{00}^{00}(f_1,f_2),C_3(f_1,f_2,f_3) = C_{000}^{000}(f_1,f_2,f_3)$
This simplifies further for $\Ng=0$
\[
\begin{aligned}
    \dot{\expval{I}} &=\Ne\left[\frac{1}{2}(\Ne - 1)\left(C_2(2k, k'-k)+C_2(k', k')\right) - \gamma^2 \right].
\end{aligned}
\]
Thus, for full initial excitation, only the $C_2[2k,k'-k]$ and $C_2[k',k']$ integrals are relevant, which are given by
\[
\begin{aligned}
    C_2[k',k'] &= \gamma^2\frac{9}{16} \iint_{-1}^{1} \diff{u}\diff{v}\me^{-\eta^2v^2}(1-u^2)(1-v^2) = \gamma^2\frac{3}{8}\frac{\left[\sqrt{\pi}(2\eta^2 - 1)\erf{\eta} + 2\eta\me^{-\eta^2}\right]}{\eta^3}\\
    C_{2}[2k,k'-k] &= \gamma^2 \frac{9}{16}\iint_{-1}^1\diff{u}\diff{v}\me^{-\eta^2u^2}\me^{-\eta^2(v-u)^2/2}(1-u^2)(1-v^2)
\end{aligned}
\]

For $\eta \rightarrow \infty$, only $C_2[k,k']$ is relevant as it contains a term $\propto \eta^{-1}$, while $C_2[2k,k'-k]$ decays faster. Therefore, for $\eta \gg 1$ we can write
\begin{equation}
    \dot I = \gamma^2\Nm\left[\frac{3\sqrt{\pi}(\Nm -1 )}{8\eta} - 1\right] 
\end{equation}
Thus, effects of particle statistics remain at finite density $\Ne/\eta=\textrm{const.}$, and the condition to obtain a burst from an initially fully excited ensemble is
\[
\frac{3\sqrt{\pi}(\Nm - 1)}{8\eta} > 1
\]

In the limit $\eta\rightarrow 0$ we can solve the integral $C_2[2k, k' - k]$ via taylor expansion of the gaussians
\[
\begin{aligned}
    C_2[k',-k'] &= \gamma^2\frac{9}{16} \iint_{-1}^{1}(1 - \eta^2u^2)(1-u^2)(1-v^2)\diff{u}\diff{v}\\
    &= \gamma^2\frac{9}{16}\left[1 - \frac{16}{45}\eta^2 + \mathcal{O}\left(\eta^4\right)\right]\\
    C_2[2k, k'-k] &= \gamma^2 \frac{9}{16} \iint_{-1}^{1} \left[1 - \eta^2(u^2 + (u-v)^2/2)\right](1-u^2)(1-v^2) \text{d}u\text{d}v\\
    &= \gamma^2\left[1 -\frac{2}{5}\eta^2 + \mathcal{O}\left(\eta^4\right)\right]\;
\end{aligned}
\]
Thus, to summarize the two limits $\eta \rightarrow 0,\eta\rightarrow \infty$ we obtain for the slope
\[
    \dot I_\text{B}(0) = 
    \begin{cases}
        \gamma^2\Nm\left[(\Nm-1)\left(1 - \dfrac{17}{45}\eta^2\right)-1\right]  + \mathcal{O}\left(\eta^4\right) \quad &\eta \rightarrow 0 \\[10pt]
         \gamma^2\Nm\left[\dfrac{3\sqrt{\pi}}{8\eta}(\Nm -1 ) - 1\right] + \mathcal{O}\left(\eta^{-2}\right)\quad &\eta \rightarrow \infty
    \end{cases}
\]

\subsection{Fermions at $T=0$}

For the initial condition $\Ng_i=0$, we obtain. If we are interested only in the $T=0$ limit, we can use $\Ne_k = 1$ if $k < \Nm$, filling only the first $\Nm$ levels (starting from $k=0$).
\[
\begin{aligned}
\dot{\langle I \rangle} &= - \Nm\gamma^2 - \sum_{m_1=0}^{\Nm-1} \Bigl[ C_{m_1m_1}^{m_1m_1}(k'-k, 2k)+ C_{m_1m_1}^{m_1m_1}(k', -k') \Bigr] \\
    &\quad + \frac{1}{2} \sum_{m_1, m_2=0}^{\Nm-1}\Bigl[C_{m_2m_1}^{m_1m_2}(k'-k, 2k) + C_{m_1m_2}^{m_1m_2}(k'-k, 2k) + C_{m_2m_1}^{m_1m_2}(k', -k') + C_{m_1m_2}^{m_1m_2}(k', -k') \Bigr]
\end{aligned}
\]
Analytical evaluation of this expression is challenging. However, numerical evaluation is easily possible. From the numerical evaluation in Fig.~\ref{fig:fig8} we obtain
\[
    \dot I_\text{F}(0) = 
    \begin{cases}
        \gamma^2\Nm(\Nm-1)\left(1 - \beta\Nm\eta^2\right)  + \mathcal{O}\left(\eta^4\right) \quad &\eta \rightarrow 0 \\[10pt]
         \gamma^2\Nm\left[\dfrac{\alpha}{\eta^3}(\Nm -1 ) - 1\right] + \mathcal{O}\left(\eta^{-4}\right)\quad &\eta \rightarrow \infty
    \end{cases}
\]
where $\alpha,\beta$ are constants with $\alpha\approx 0.4,$ and  $\mathcal{O}(\beta) = 1$.

\end{document}